\documentclass[twocolumn,showpacs,showkeys,preprintnumbers,amsmath,amssymb,floatfix]{revtex4}

\usepackage{graphicx}
\usepackage{dcolumn}
\usepackage{bm}
\usepackage{rotating}

\newcommand{\bea}{\begin{eqnarray}}
\newcommand{\eea}{\end{eqnarray}}
\newcommand{\etal}{{\it et al.\ }}
\newcommand{\jpi}{$J\pi$ }
\newcommand{\nga}{(n,$\gamma$) }
\newcommand{\ngax}{(n,$\gamma$)}
\newcommand{\nf}{(n,f) }
\newcommand{\we}{Weisskopf-Ewing }
\newcommand{\cn}{compound nucleus }
\newcommand{\uthree}{$^{233}$U }
\newcommand{\ufour}{$^{234}$U }
\newcommand{\ufive}{$^{235}$U }
\newcommand{\usix}{$^{236}$U }
\newcommand{\uthreex}{$^{233}$U}
\newcommand{\ufourx}{$^{234}$U}
\newcommand{\ufivex}{$^{235}$U}
\newcommand{\usixx}{$^{236}$U}
\newcommand{\gdfive}{$^{155}$Gd }
\newcommand{\gdsix}{$^{156}$Gd }
\newcommand{\gdseven}{$^{157}$Gd }
\newcommand{\gdfivex}{$^{155}$Gd}
\newcommand{\gdsixx}{$^{156}$Gd}
\newcommand{\gdsevenx}{$^{157}$Gd}
\newcommand{\idest}{{\it i.e.\ }}

\begin{document}


\title{Cross sections for neutron capture from surrogate measurements:\\
An examination of Weisskopf-Ewing and ratio approximations}

\author{Jutta E. Escher}\email{escher1@llnl.gov}
\author{Frank S. Dietrich}
\affiliation{
Lawrence Livermore National Laboratory, Livermore, CA 94550
}

\date{\today}

\begin{abstract}
Motivated by the renewed interest in the surrogate nuclear reactions approach, an indirect method for determining compound-nuclear reaction cross sections, the prospects for determining \nga cross sections for deformed rare-earth and actinide nuclei are investigated.
A nuclear-reaction model is employed to simulate physical quantities that are typically measured in surrogate experiments and used to assess the validity of the \we and ratio approximations, which are typically employed in the analysis of surrogate reactions.  
The expected accuracy of \nga cross sections extracted from typical surrogate measurements is discussed and limitations of the approximate methods are illustrated. Suggestions for moving beyond presently-employed approximations are made.
\end{abstract}
 

\maketitle


\section{Introduction}
\label{sec_intro}

Compound-nuclear reactions play an important role in many areas of basic and applied nuclear science.  The production of heavy elements in various astrophysical environments, for example, involves compound reactions and the resulting observable abundance patterns depend, sometimes very sensitively, on the associated reaction cross sections~\cite{Kaeppeler:06,surman:09,Beun:09}.  Similarly, a proper description of nuclear fuel cycles for energy applications requires data on various types of compound reactions~\cite{afci:06}.

Often the cross section needed for a particular application cannot be measured directly since the relevant energy region is inaccessible or the target is too short-lived.  To overcome the experimental limitations, indirect methods, such as the {\em Surrogate Nuclear Reactions} approach, have to be developed.  In this approach the compound nucleus ($B^*$) occurring in the reaction of interest ($a+A \rightarrow B^* \rightarrow c+C$) is produced via an alternative, ``surrogate'' reaction ($d+D \rightarrow B^* + b$) that involves a projectile-target combination ($d + D$) that is experimentally more accessible.  The measured compound-nuclear decay probabilities can then be combined with calculated formation cross sections for the compound nucleus in the desired reaction to yield the relevant reaction cross section.

Originally introduced in the 1970s~\cite{Cramer:70b,Britt:79}, the surrogate approach has recently received renewed attention~\cite{Younes:03a,Younes:03b,Petit:04,Boyer:06,Burke:06a,EscherDietrich:06PRC,Jurado:07cnr,Escher:07ND}. 
A large number of surrogate experiments aimed at obtaining (n,f) cross sections has been carried out over the years, whereas few experiments have been designed to determine \nga cross sections.  Still fewer experiments have attempted to provide information about the charged-particle or two-neutron exit channels.  In this paper, we focus on the prospects of determining \nga cross sections from surrogate experiments.  
In particular, we examine the validity of commonly-employed approximation schemes which ignore the ``spin-parity mismatch,'' \idest the difference in the spin-parity distributions of the compound nuclei produced in the desired and surrogate reactions, respectively.
An earlier theoretical study of zirconium isotopes~\cite{Forssen:07} demonstrated that the probabilty for decay of a compound nucleus via gamma emission depends very sensitively on its spin-parity population in this mass region, in particular for isotopes near closed shells. In the present paper, we examine the prospects for extracting \nga cross sections for deformed rare-earth and actinide nuclei.  The higher level densities in these mass regions are expected to reduce the sensitivity of the $\gamma$-decay probabilities to the compound-nuclear spin-parity distribution.  
The work presented here complements our previous study of the validity of surrogate reactions to determine \nf cross sections~\cite{EscherDietrich:06PRC}.

In the next section (Section~\ref{sec_concepts}), we summarize the surrogate idea and establish our notation. 
In Section~\ref{sec_challenges}, we describe the challenges that particularly affect the determination of \nga reaction cross sections from surrogate measurements. In Section~\ref{sec_method}, we describe the simulations used to study the approximation schemes.  
The results are discussed in Sections~\ref{sec_actinides} and \ref{sec_rareearths}, for the actinide and rare-earth cases, respectively, and  conclusions are given in Section~\ref{sec_conclusions}.  The appendix contains information on the Flap 2.2 optical-model potential for the actinide region, which was used here, as well as in various recent applications of the surrogate method.


\section{Surrogate Approaches}
\label{sec_concepts}

This section establishes the notation employed in this paper; additional details about the surrogate formalism can be found in Ref.~\cite{EscherDietrich:06PRC}.

\subsection{The Surrogate Idea}
\label{subsec_concepts_surrogates}

Compound-nuclear reactions are properly described in the Hauser-Feshbach formalism~\cite{HauserFeshbach:52}, which takes account of the  conservation of angular momentum $J$ and parity $\pi$.  The cross section for the ``desired'' reaction with entrance and exit channels $\alpha$ $=a+A$ and $\chi$ $=c+C$, respectively, is written as:
\begin{eqnarray}
\sigma_{\alpha \chi}(E_{a}) &=& \sum_{J,\pi}  \sigma_{\alpha}^{CN}(E_{ex},J,\pi) \;\; G_{\chi}^{CN}(E_{ex},J,\pi) \; .
\label{eq:DesReact}
\end {eqnarray}
\noindent
The excitation energy $E_{ex}$ of the compound nucleus, $B^*$, is related to the center-of-mass energy $E_a$ in the entrance channel via the energy needed for separating $a$ from $B$: $E_a=E-S_a(B)$.  
The objective of the surrogate method is to experimentally determine or constrain the decay probabilities $G_{\chi}^{CN}(E_{ex},J,\pi)$, which are often difficult to calculate accurately.

In a surrogate experiment, the compound nucleus $B^*$ is produced via an alternative (``surrogate"), direct reaction
$d+D \rightarrow b+B^*$ and the decay of $B^*$ is observed in coincidence with the outgoing particle $b$.
The probability for forming $B^*$ in the surrogate reaction (with specific values for $E_{ex}$, $J$, $\pi$) is $F_{\delta}^{CN}(E_{ex},J,\pi)$,
where $\delta$ refers to the entrance channel reaction $D(d,b)$. The quantity
\begin{equation}
P_{\delta\chi}(E_{ex}) = \sum_{J,\pi} F_{\delta}^{CN}(E_{ex},J,\pi) \;\; G_{\chi}^{CN}(E_{ex},J,\pi) \; ,
\label{Eq:SurReact}
\end{equation}
which gives the probability that the compound nucleus $B^*$ was formed with energy $E_{ex}$
and decayed into channel $\chi$, can be obtained experimentally, by measuring $N_{\delta}$, the total number of surrogate events, and $N_{\delta\chi}$, the number of coincidences between the direct-reaction particle and the observable that identifies the relevant exit channel:
\begin{equation}
P^{exp}_{\delta\chi}(E_{ex}) = \frac{N_{\delta\chi}}{N_{\delta}} \; .
\label{Eq:CoincProb}
\end{equation}
For simplicity, we have omitted here the efficiencies for detecting the outgoing direct-reaction particle $b$ and the exit channel $\chi$, as well as the angular dependence of both the desired and surrogate reactions. 

To determine the desired cross section from a surrogate measurement, one can pursue the following strategies:

\paragraph*{I. Ideal Approach.}
Ideally, one calculates the spin-parity distribution, $F_{\delta}^{CN}(E_{ex},J,\pi)$, in Eq.~\ref{Eq:SurReact} from a suitable theory that describes the formation of the compound nucleus following the direct reaction $d+D \rightarrow b+B^*$.
Given a reliable prediction of the quantities $F_{\delta}^{CN}(E_{ex},J,\pi)$, and a sufficient range of experimental coincidence data $P_{\delta\chi}(E_{ex})$ (for a range of energies and angles of the outgoing particle $b$, and possibly for various exit channels), it might be possible to extract the $G_{\chi}^{CN}(E_{ex},J,\pi)$ which can then be used to calculate the desired cross section, Eq.~\ref{eq:DesReact}.  At this time, this idealized approach has not been implemented since a combination of possible reaction mechanisms, predicted $F_{\delta}^{CN}$, and experimental data has not been available to unambiguously extract useful branching ratios.

\paragraph*{II. Modeling Approach.}
More realistically, the decay of the compound nucleus is modeled in a Hauser-Feshbach-type calculation that makes use of independently available (but typically incomplete) nuclear structure information.  The $G_{\chi}^{CN}(E,J,\pi)$ obtained from such modeling are combined with calculated $F_{\delta}^{CN}(E_{ex},J,\pi)$ to yield a prediction for $P_{\delta\chi}(E_{ex})$.  Fitting the latter to surrogate data provides further constraints on the $G_{\chi}^{CN}$ which can then be employed in the calculation of the desired cross section.
Steps towards developing this modeling approach were taken by Andersen \etal \cite{Andersen:70}, Back \etal \cite{Back:74a}, and Younes and Britt~\cite{Younes:03a,Younes:03b} for measurements designed to yield (n,f) cross sections. 

\paragraph*{III. Approximations.}
A large majority of the surrogate applications to date has relied on invoking approximations, such as the \we limit of the Hauser-Feshbach theory, which treats the $G_{\chi}^{CN}(E,J,\pi)$ as independent of $J\pi$, or the surrogate ratio method; these are further described below and in Ref.~\cite{EscherDietrich:06PRC}.  

\paragraph*{IV. ``Serendipitous'' (``Matching'') Approach.}
A primary challenge for the surrogate approach lies in accounting for the spin-parity mismatch between the desired and surrogate reactions. When it is possible to identify a surrogate reaction ({\idest a reaction mechanism, projectile-target combination, beam energy, outgoing-particle angle) that approximately reproduces the spin-parity distribution of the desired reaction, 
\bea
\lefteqn{F_{\delta}^{CN}(E_{ex},J,\pi)} \nonumber \\
&& \approx F_{\alpha}^{CN}(E_{ex},J,\pi) \equiv \frac{\sigma_{\alpha}^{CN}(E_{ex},J,\pi)}
{\sum_{J^{\prime},\pi^{\prime}}  \sigma_{\alpha}^{CN}(E_{ex},J^{\prime},\pi^{\prime}) } ,
\label{Eq:Lucky}
\eea
where $F_{\alpha}^{CN}$ is the compound-nuclear \jpi population in the desired reaction, 
the situation simplifes greatly, as in this limit, we find:
\bea
\sigma_{\alpha \chi}(E_{a}) &\approx& \sigma_{\alpha}^{CN}(E_{ex}) \times P_{\delta\chi}^{exp}(E_{ex}),
\eea
where the cross section for forming the compound nucleus at energy $E_{ex}$,
\begin{eqnarray}
\sigma^{CN}_{\alpha}(E_{ex}) &\equiv&  \sum_{J\pi} \sigma^{CN}_{\alpha} (E_{ex},J,\pi) ,
\label{Eq:WEReactXSec}
\end{eqnarray}
is calculated using a suitable optical potential, and $P_{\delta\chi}^{exp}(E_{ex})$ is determined from the experiment.
While it is sometimes argued that a given surrogate experiment approximately satisfies Eq.~\ref{Eq:Lucky}, there has not been sufficient evidence to support such claims.


\subsection{The Weisskopf-Ewing limit of the Hauser-Feshbach description}
\label{subsec_concepts_we}

In the Weisskopf-Ewing limit of the Hauser-Feshbach theory the branching ratios are independent of angular momentum
and parity, $G_{\chi}^{CN}(E_{ex},J,\pi)$ $\rightarrow$ ${\cal G}^{CN}_{\chi}(E_{ex})$, and the cross section expression for the desired reaction becomes
\begin{eqnarray}
\sigma^{WE}_{\alpha \chi}(E_a) &=& \sigma^{CN}_{\alpha}(E_{ex}) \;  {\cal G}^{CN}_{\chi}(E_{ex}) \; ,
\label{Eq:WELimit}
\end{eqnarray}
with $\sigma^{CN}_{\alpha}(E_{ex})$ as defined in Eq.~\ref{Eq:WEReactXSec}. 
The Weiskopf-Ewing approximation greatly simplifies the application of the surrogate method: The branching ratios ${\cal G}^{CN}_{\chi}$ can be directly obtained from the measured coincidence probabilities $P_{\delta\chi}$ (since $\sum_{J\pi} F^{CN}_{\delta} (E_{ex},J,\pi)$ $= 1$),
\begin{eqnarray}
P_{\delta\chi}(E_{ex}) &=& {\cal G}^{CN}_{\chi}(E_{ex}) \; ,
\label{Eq:WECoincProb}
\end{eqnarray}
and the desired cross section can be written as
\bea
\sigma^{CN}_{\alpha}(E_{ex}) &=&  \sigma^{CN}_{\alpha} (E_{ex}) P_{\delta\chi}(E_{ex}) \; ,
\label{Eq:WElimitXSec}
\eea
\idest calculating the direct-reaction probabilities $F_{\delta}^{CN}(E_{ex},J,\pi)$ or modeling the compound-nuclear decay is not required. Most applications of the surrogate method so far have made use of the \we approximation.

It can be formally shown that the Hauser-Feshbach theory reduces to the \we limit when a set of conditions is satisfied~\cite{Gadioli:92,Dietrich:04}: width-fluctuation correlations have to be negligible, the decay of the \cn to discrete states of nuclei in the various exit channels has to be small, and the level densities in the decay channels have to possess a particular dependence on the spins of the states in the residual nuclei, namely $\rho \propto 2J+1$. Since most of these conditions tend to be satisfied at higher compound-nuclear energies, it is often assumed that the \we approximation can be employed above a certain bombarding energy, e.g. above 1-2 MeV in neutron-induced reactions.  In reality, the situation is more complex and it is not {\em a priori} clear whether the \we approximation is valid in a particular energy regime.  
An obvious difficulty with the \we limit is the fact that the spin dependence of realistic level densities is approximated by 
\bea
\rho(J) &\propto& (2J+1) e^{-(2J+\frac{1}{2})^2/2\sigma^2} \; ,
\label{Eq:levDens}
\eea
where $\sigma$ is the spin-cutoff parameter~\cite{GilbertCameron}.  Thus the necessary condition on the spin dependence for the \we formula is satisfied only for spins significantly lower than the spin cutoff parameter.  A breakdown of the \we approximation may be anticipated when $J \geq \sigma$.  


\subsection{Surrogate analyses using ratios}
\label{sec_concepts_ratio}

The goal of the {\em Surrogate Ratio approach}~\cite{EscherDietrich:06PRC,Burke:06a} is to experimentally determine the ratio 
\begin{eqnarray}
R(E)&=& \frac{\sigma^{CN1}_{\alpha_1 \chi_1}(E)}{\sigma^{CN2}_{\alpha_2 \chi_2}(E)}
\label{Eq:RatioDef}
\end{eqnarray}
of the cross sections of two compound-nucleus reactions,
$a_1 + A_1 \rightarrow B_1^* \rightarrow c_1 + C_1$ and $a_2 + A_2 \rightarrow B_2^* \rightarrow c_2 + C_2$.
One of the cross sections, say $\sigma_{\alpha_2 \chi_2}(E)$, needs to be known, and the other $\sigma_{\alpha_1 \chi_1}(E)$ is 
extracted from the ratio.  In the Weisskopf-Ewing limit,
\begin{eqnarray}
R(E)&=& \frac{\sigma^{CN1}_{\alpha_1}(E) \;  {\cal G}^{CN1}_{\chi_1}(E)}{\sigma^{CN2}_{\alpha_2}(E) \;  {\cal G}^{CN2}_{\chi_2}(E)} \; ,
\label{Eq:RatioWELimit}
\end{eqnarray}
with branching ratios ${\cal G}^{CN}_{\chi}(E)$ that are independent of the $J\pi$ populations of the compound nuclei.
For most cases of interest the compound-nucleus formation cross sections $\sigma^{CN1}_{\alpha_1}$ and
$\sigma^{CN2}_{\alpha_2}$ can be calculated adequately by using an optical model.

To determine ${\cal G}^{CN1}_{\chi_1}(E)$ / ${\cal G}^{CN2}_{\chi_2}(E)$, two experiments are carried out that create the compound nuclei, $CN1$ and $CN2$, respectively.  For each experiment, the number of coincidence events, $N^{CN1}_{\delta_1\chi_1}$ and $N^{CN2}_{\delta_2\chi_2}$, is measured.  The ratio of the branching ratios is given by
\begin{eqnarray}
\frac{{\cal G}^{CN1}_{\chi_1}(E)} {{\cal G}^{CN2}_{\chi_2}(E)} 
\!\!&=&\!\! \frac{P^{CN1}_{\delta_1\chi_1}(E)} {P^{CN2}_{\delta_2\chi_2}(E)}
\! = \!  \frac{N^{CN1}_{\delta_1\chi_1}(E) } {N^{CN2}_{\delta_2\chi_2}(E) } 
\frac{N^{CN2}_{\delta_2}(E)}{N^{CN1}_{\delta_1}(E)} .
\label{Eq:CoincRatio}
\end{eqnarray}
\noindent
The experimental conditions are adjusted such that the relative number of reaction events, $N^{CN1}_{\delta_1}/$  $N^{CN2}_{\delta_2}$, 
can be determined from the relative beam intensities, target thickness, and livetimes of the two experiments.  
$R(E)$ then becomes:
\begin{eqnarray}
R^{exp}(E)&=& \frac{\sigma^{CN1}_{\alpha_1}(E) \; N^{CN1}_{\delta_1\chi_1}(E)} {\sigma^{CN2}_{\alpha_2}(E) \; N^{CN2}_{\delta_2\chi_2}(E)} \;,
\label{Eq:ExpRatio}
\end{eqnarray}
where we have set $N^{CN1}_{\delta_1}/$  $N^{CN2}_{\delta_2}$ = 1
to simplify the notation.  
The definition of the energy $E$ in Eqs.~\ref{Eq:RatioDef}--\ref{Eq:ExpRatio} is discussed below (see ``Energy matching'').

Different variants of the surrogate ratio approach can be considered, depending on the entrance and exit channels of interest:

\paragraph*{External Surrogate Ratio (ESR) Approach.}  This is the most-widely employed variant of the ratio approach~\cite{Burke:06a,Plettner:05,EscherDietrich:06PRC,Lyles:07PRC,Lesher:09}. The cross sections in $R(E) =$ $\sigma^{CN1}_{\alpha_1 \chi_1} / \sigma^{CN2}_{\alpha_2 \chi_2}$ refer to two reactions with the same type of entrance channel, $\alpha_1 =$ $\alpha_2$, and the same type of exit channel, $\chi_1 =$ $\chi_2$, but different compound nuclei, $CN1 \ne CN2$.  The surrogate measurements involve identical entrance channels, $\delta_1 =$ $\delta_2$, and exit channels.  
In Ref.~\cite{Burke:06a}, for instance, the ratio $\sigma[^{237}$U(n,f)] / $\sigma[^{235}$U(n,f)] was determined from measurements of $P[^{238}$U$(\alpha,\alpha^{\prime}f)]$ / $P[^{236}$U($\alpha,\alpha^{\prime}f)]$, where $\alpha$ and $\alpha^{\prime}$ refer to alpha particles, not channels. The entrance and exit channel were $\alpha_1 =$ $\alpha_2 =$ $n+target$, and $\chi_1 =$ $\chi_2 =$ $fission$.
To determine $R(E)$, it is necessary to take into account the ratio $\sigma^{CN1}_{\alpha_1} / \sigma^{CN2}_{\alpha_2}$.  In many applications of the ESR method, this ratio has simply been set to one, but this is not necessarily a good approximation (see the discussion on energy matching below).

\paragraph*{Internal Surrogate Ratio (ISR) Approach.}  In  this variant~\cite{Bernstein:06TR_IR, Bernstein:05,Allmond:09}, the compound nuclei created in the two reactions of interest are identical, $CN1 = CN2$, the entrance channels are identical, $\alpha_1 =$ $\alpha_2$, but the decay channels differ in type, $\chi_1 \neq$ $\chi_2$. The surrogate measurement employs {\em one} projectile-target combination, $\delta_1 =$ $\delta_2$.  
For example, in Ref.~\cite{Allmond:09} the ratio $\sigma[^{235}$U(n,$\gamma$)] / $\sigma[^{235}$U(n,f)] was determined from a measurement of $P[^{235}$U(d,p$\gamma$)] / $P[^{235}$U(d,pf)], {\it i.e.\ } $\alpha_1 =$ $\alpha_2 =$ n + $^{235}$U, but $\chi_1 \neq$ $\chi_2$.
Since the entrance channels and compound nuclei involved are identical, one can set $\sigma^{CN1}_{\alpha_1} / \sigma^{CN2}_{\alpha_2} =$ 1, provided the decay probabilities in Eq.~\ref{Eq:CoincRatio} are compared at the proper energies (see ``Energy matching''). 

\paragraph*{Other variants.}  In Ref.~\cite{Nayak:08}, surrogate $^{232}$Th($^6$Li,$\alpha$)$^{234}$Pa and $^{232}$Th($^6$Li,d)$^{236}$U reactions were used to infer information on the cross section ratio $\sigma$[$^{233}$Pa(n,f)] / $\sigma$[$^{235}$U(n,f)]. The desired and reference reactions were both of the same type, namely (n,f), but two different surrogate mechanisms were employed for producing the compound nuclei, namely ($^6$Li,$\alpha$) and ($^6$Li,d).

\subsubsection*{Energy matching in the ratio approach}
\label{sec_ratio_ematch}

In a surrogate ratio analysis, the choice of energy variable at which the data sets are compared (Eqs.~\ref{Eq:RatioDef}--\ref{Eq:ExpRatio}) introduces a subtle but important issue that can affect the results, even when the \we approximation is valid. 
The comparison of the cross sections for the reactions $a_1 + A_1 \rightarrow B_1^* \rightarrow c_1 + C_1$ (numerator) and
$a_2 + A_2 \rightarrow B_2^* \rightarrow c_2 + C_2$ (denominator) can be made either {\em at the same projectile energy} $E_a$ or 
{\em at the same excitation energy} $E_{ex}$.  In a compound-nucleus reaction, those two energies are related via the separation energy $S_a$ of the particle $a$ in $B^*$:  $E_{ex}=S_a+E_a$.  

The energy-dependence of $\sigma^{CN}_{\alpha}$ is most naturally characterized by the kinetic energy of the projectile, $E_a$.  When the cross sections in Eq.~\ref{Eq:RatioDef} are compared at the same projectile energy, the ratio  $\sigma^{CN}_{\alpha_1}$ / $\sigma^{CN}_{\alpha_2}$ can sometimes be approximately set to one for the relevant energy range. This is convenient, as the calculation of two formation cross sections and the associated uncertainties can be avoided in this case. For the ESR method, this approximation is likely to be valid if {\em one} projectile type is considered,  $a_1 = a_2$, hitting targets that are structurally similar (deformation, level structure), such as $^{233}$U and $^{235}$U.
For the ISR method, this ratio is by definition one, provided the energies are matched at $E_a$.  In Ref.~\cite{Nayak:08} (n,f) reactions on Pa and U targets were compared, so the ratio had to be explicitly calculated.

Matching the energies of numerator and denominator in Eq.~\ref{Eq:RatioDef} at the projectile energy, on the other hand, may introduce experimental challenges:  For a given projectile energy, $E_{a_1}=$ $E_{a_2}$, differences in the separation energies, $S_{a_1}$ and $S_{a_2}$, lead to different excitation energies in the compound nuclei, $B_1^*$ and $B_2^*$, respectively, and thus to different kinetic energies for the outgoing direct-reaction particles $b_1$ and $b_2$.  The difference in the (energy-dependent) efficiencies for detecting these particles needs to be accounted for explicitly~\cite{Scielzo:09ip}. 


\section{Challenges for surrogate measurements of $(n,\gamma)$ cross sections}
\label{sec_challenges}

For \nf reactions, the spin mismatch between the surrogate and desired reactions was found to primarily affect the accuracy of the extracted cross sections at low energies ($E_n <$ 1 MeV), and, to a lesser extent, at the onset of first and second-chance fission~\cite{EscherDietrich:06PRC}.
Since the energy region of interest to many applications that require neutron-capture cross sections lies below about 1 MeV, accounting for this mismatch is expected to be very important.  To investigate this, we calculated the $\gamma$-decay probabilities $G_{\gamma}^{CN}(E,J,\pi)$ for the compound nucleus \usixx.  
We carried out a standard Hauser-Feshbach calculation for the \ufivex(n,f) and \ufivex\nga cross sections, fitted parameters to measured cross sections, and extracted the individual $\gamma$-decay probabilities.  
Selected $\gamma$ branching ratios $G_{\gamma}^{CN}(E,J,\pi)$ are shown in Fig.~\ref{fig:gamProbsU236} for \usix excitation energies $E_{ex} =$ 6.55-10.5 MeV, which corresponds to neutron energies $E_n =$ 0-4 MeV.

\begin{figure}[htb]
\centering
\resizebox{1.0\columnwidth}{!}{
    \includegraphics[viewport=80 15 525 730,clip,angle=270]{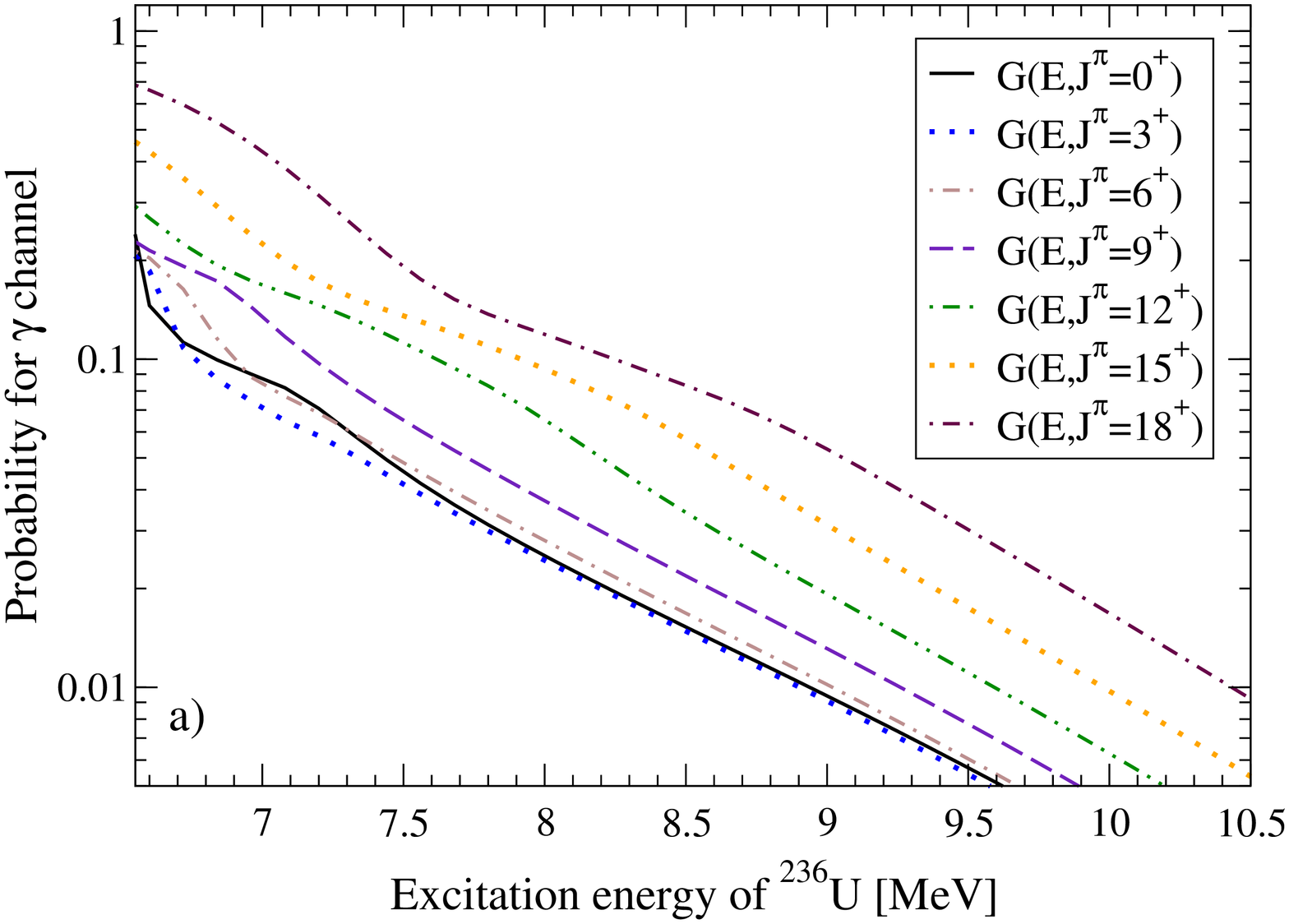}
}
\hfill
\resizebox{1.0\columnwidth}{!}{
    \includegraphics[viewport=80 15 595 730,clip,angle=270]{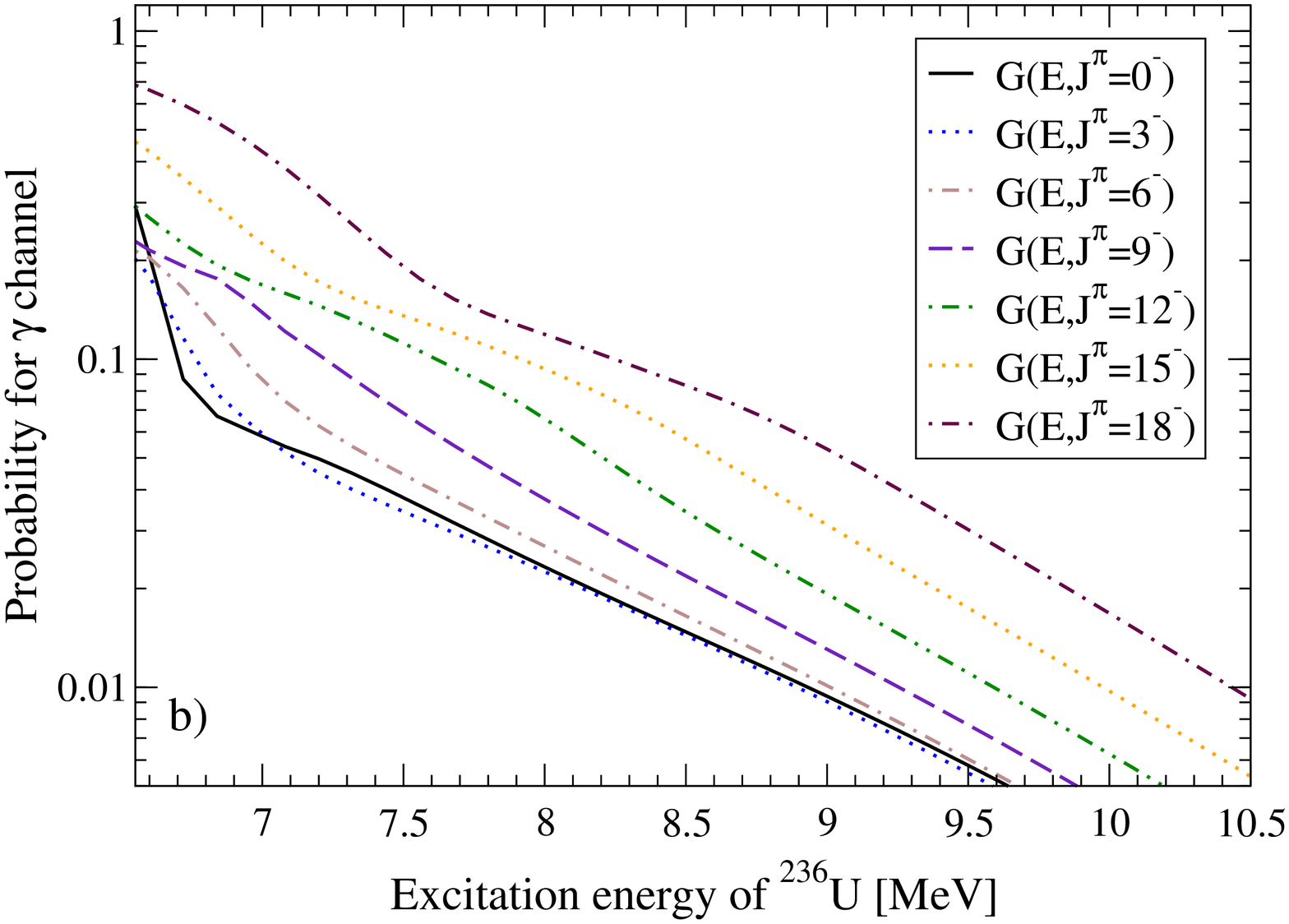}    
}
\caption[Calculated $\gamma$-decay probabilities $G^{CN}_{\gamma}(E,J,\pi)$, for \usix]{(Color online) Calculated $\gamma$-decay probabilities $G^{CN}_{\gamma}(E,J,\pi)$, for \usixx. Shown is the probability that the compound nucleus, when produced with a specific \jpi combination, decays via the $\gamma$ channel.  Positive-parity decay probabilities are shown in panel a), and negative-parity decay probabilities are shown in panel b).
}
\label{fig:gamProbsU236}
\end{figure}

While the $G_{\gamma}^{CN}(E,J,\pi)$ for $J = 0-6$ are very similar to each other for $E_n >$ 1 MeV, they differ more significantly from each other below 1 MeV.  For energies above $E_n \approx$ 1.5 MeV, all branching ratios exhibit roughly the same energy dependence, but the $G_{\gamma}^{CN}(E,J,\pi)$ associated with the higher angular-momentum values  $J = 9,12$ differ from those for $J = 0,3$ by factors ranging from 0.5 to 3; for $J = 15,18$ the difference is a factor of 3-6.
Given the fact that the compound nucleus can exhibit spin-parity distributions peaked at various ranges of spins, depending on the reaction that produces it, we expect, based on these calculations, the cross sections obtained in the \we approximation to be limited in accuracy.  This issue is investigated in detail in Section~\ref{sec_actinides}.

The behavior of the branching ratios $G_{\gamma}^{CN}(E,J,\pi)$ is governed by the competition of fission, neutron emission, and $\gamma$ decay. Fission competes with $\gamma$ emission below the neutron separation threshold, resulting in $G_{\gamma}^{CN}(E,J,\pi) < 1$ at $E_{ex} =7.55$ MeV ($E_n =0$).
The rough equality of the $\gamma$ probabilities for $J \leq 6$ and the significant increase in the probabilities for larger $J$ values is a consequence of the breakdown of the \we approximation due to the spin-cutoff effect in the level densities as discussed earlier (see Eq.~\ref{Eq:levDens}). Similar behavior was observed for the fission probabilities $G_{fission}^{CN}(E,J,\pi)$, see Ref.~\cite{EscherDietrich:06PRC}, Figs. 8 and 9
\footnote{Values of the spin-cutoff parameter are not well determined from experiment, and theoretical models vary in their predictions according to structural details such as the moment of inertia.  Thus the exact values of the probabilities at high spins are highly uncertain, both for the $\gamma$ probabilities discussed here and for the fission probabilities estimated in Ref.~\cite{EscherDietrich:06PRC}.
}.
An increased probability for $^{236}$U states with larger $J$ values to decay via $\gamma$ emission is not surprising, as s-wave neutron emission from these states is hindered at low energies due to angular-momentum selection rules, but also not immediately obvious, as the fission channel has to be considered. The situation is clearer for the rare earth cases discussed in Sec.~\ref{sec_rareearths}, where neutron and $\gamma$ emission are the only significant decay modes. 

An additional challenge that has to be addressed when determining \nga cross sections from surrogate experiments lies in the identification of the $\gamma$ exit channel. 
The outgoing direct-reaction particle $b$ has to be detected in coincidence with an observable that identifies the $\gamma$-emission decay channel. In current applications, this is typically accomplished by gating on coincidences between particle $b$ and individual $\gamma$ rays that are characteristic of transitions between low-lying levels of the decaying nucleus.
The experiments measure the yields of individual gammas in the $\gamma$ cascade rather than the quantity that is wanted, which is the sum of all cascades.  The fraction of the cascade that proceeds through a particular $\gamma$ transition depends on the spin-parity distribution of the decaying compound nucleus, which complicates the interpretation of the experiment.
This differs from the fission case, in which observation of fission fragments provides a direct measure of the desired quantity.  

The effect is illustrated in Fig.~\ref{fig:gambrU} for the decay of the compound nucleus $^{236}$U, formed in the $n + ^{235}$U and $n+ ^{235m}$U channels, respectively.  Here, $^{235m}$U refers to the first excited state of \ufive at $E_{ex} =$ 77 eV.  The plot shows the ratio of the calculated intensity of a particular $\gamma$ transition to the total intensity of $\gamma$ cascades that eventually reach the ground state of $^{236}$U. Internal conversion, which affects the $\gamma$ yield measured in any experiment that focuses on $\gamma$ cascades has not been considered here, but has to be accounted for in actual measurements; that is, each curve in Fig.~\ref{fig:gambrU} actually represents the complete decay rate of the state, not just the $\gamma$-emitting part.
Both panels of Fig.~\ref{fig:gambrU} show relative yields for the decay of the compound nucleus $^{236}$U as a function of energy.  Apart from the intensities for the 2$^+$ $\rightarrow$ 0$^+$ transition, the yields shown in the two panels of the figure are very different from each other.  This difference can be attributed to a difference in the \jpi distribution in the decaying compound nucleus.  The compound nucleus  $^{236}$U associated with panel a) has a spin distribution that is peaked at higher angular-momentum values than that for the compound nucleus $^{236}$U associated with panel b) (cf.\  also Fig.~\ref{fig:jpi_n235U}).  The former nucleus was produced in a reaction in which a neutron was absorbed by the \jpi=$7/2^-$ ground state of $^{235}$U, while the latter was produced in a reaction involving the first excited state, $^{235m}$U, which has angular momentum and parity \jpi=$1/2^+$.  The energy difference between these two target states is very small, 77~eV, thus the only significant difference between the compound nuclei $^{236}$U produced in these reactions is the spin-parity population.  

\begin{figure}[htb]
\centering
\resizebox{0.97\columnwidth}{!}{
     \includegraphics[viewport=80 15 530 720,clip,angle=270]{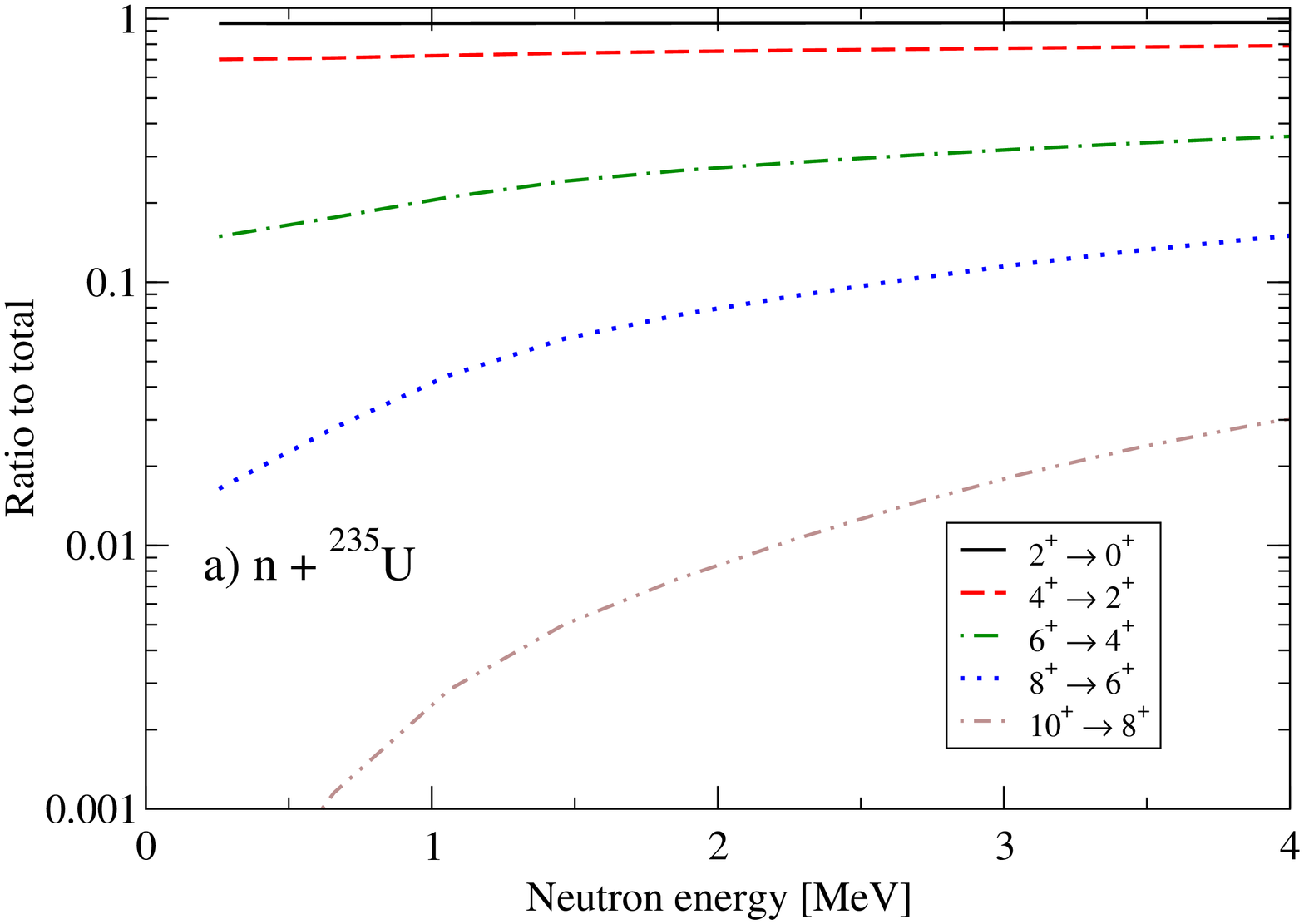}
}
\hfill
\resizebox{0.97\columnwidth}{!}{
    \includegraphics[viewport=80 15 580 720,clip,angle=270]{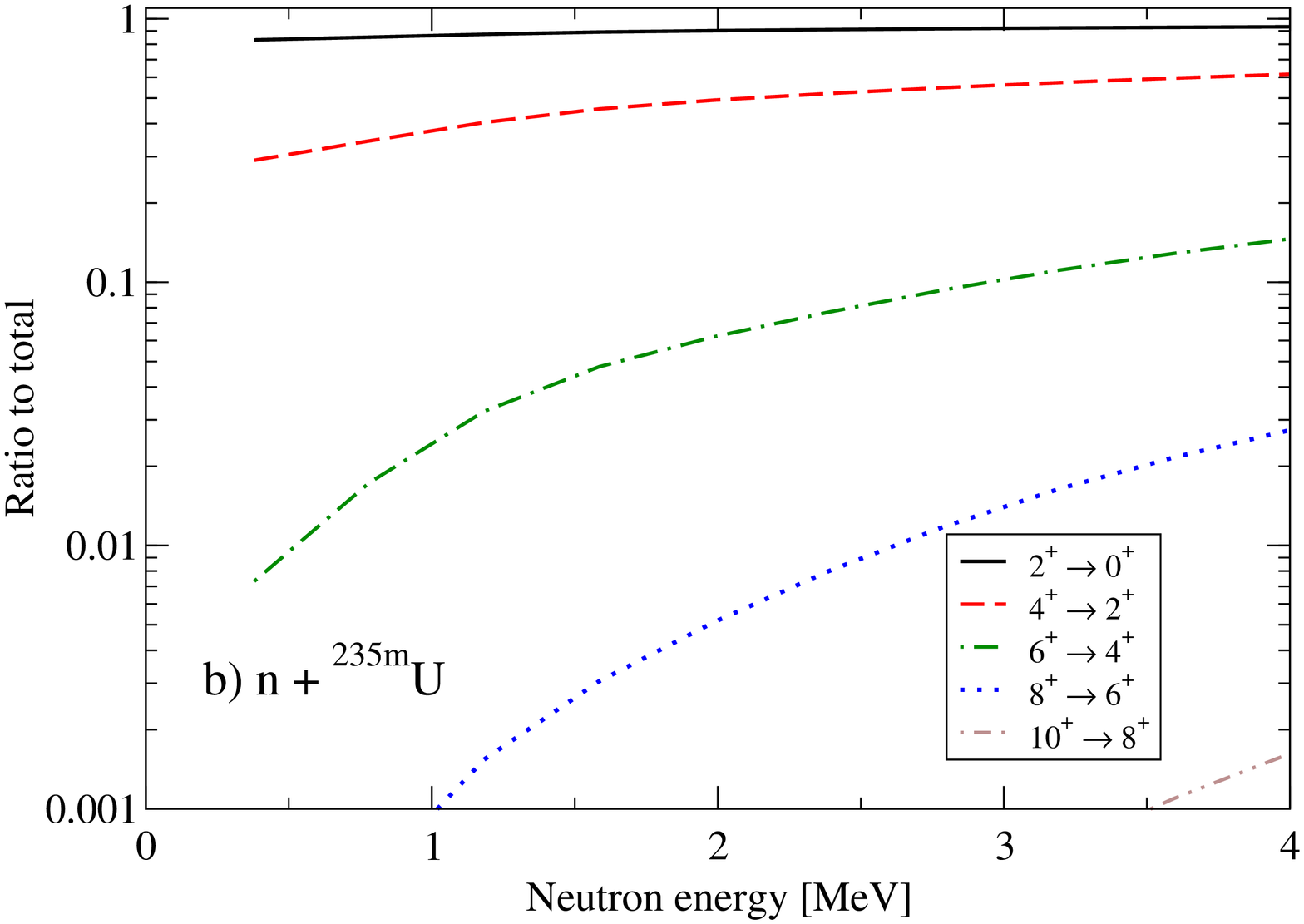}
}
\caption[Calculated yields of $\gamma$ rays for transitions in the ground-state band of $^{236}$U]{(Color online) Ratio of yields of various $\gamma$ rays for transitions in the ground-state band of $^{236}$U to the total production of $^{236}$U.  Panel a) shows results for the decay of $^{236}$U following its production in the $n$ +$^{235}$U channel, while the panel b) is for the $n$ +$^{235m}$U channel.  The associated \jpi distributions are shown in Fig.~\ref{fig:jpi_n235U}.
}
\label{fig:gambrU}
\end{figure}

It is clear that the ratios of the individual $\gamma$-ray yields to the total yield of all $\gamma$ cascades (`ratios-to-total') are highly dependent on the spin-parity distribution for all of the transitions except the $2^+\rightarrow0^+$.  This transition is dominated by internal conversion and is therefore very difficult to measure with the $\gamma$ detection techniques used in current surrogate experiments~\cite{Allmond:09,Scielzo:09ip,Hatarik:07cnr,Hatarik:10}.
Overall, it is evident that the compound-nucleus spin distribution has a significant influence on the observed quantities and thus on cross sections that are extracted if these aspects are not properly modeled.

\begin{figure*}[htb]
\centering
\resizebox{1.0\columnwidth}{!}{
    \includegraphics[viewport=30 0 590 800,clip,angle=0]{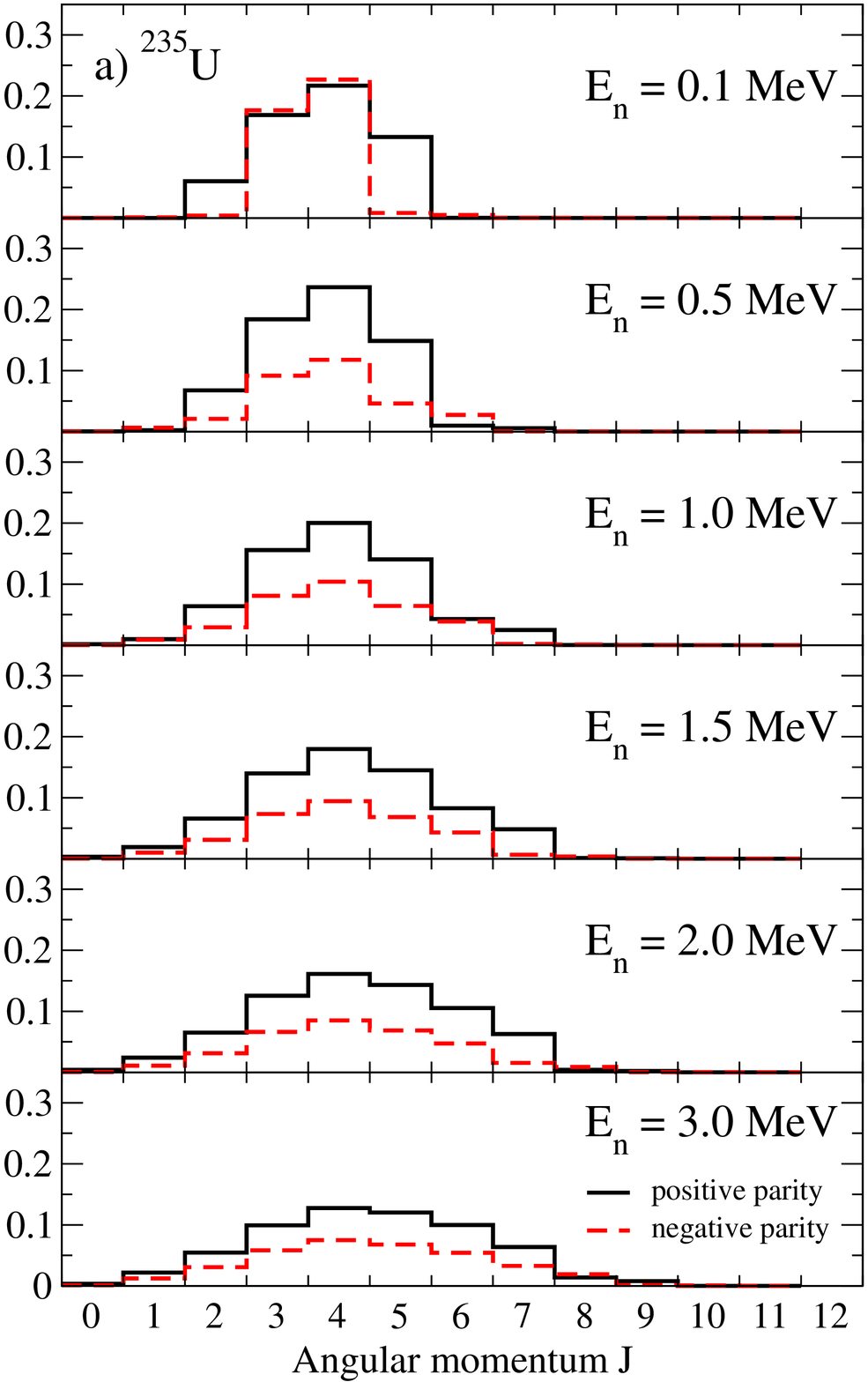}
}
\hfill
\resizebox{1.0\columnwidth}{!}{
    \includegraphics[viewport=30 0 590 800,clip,angle=0]{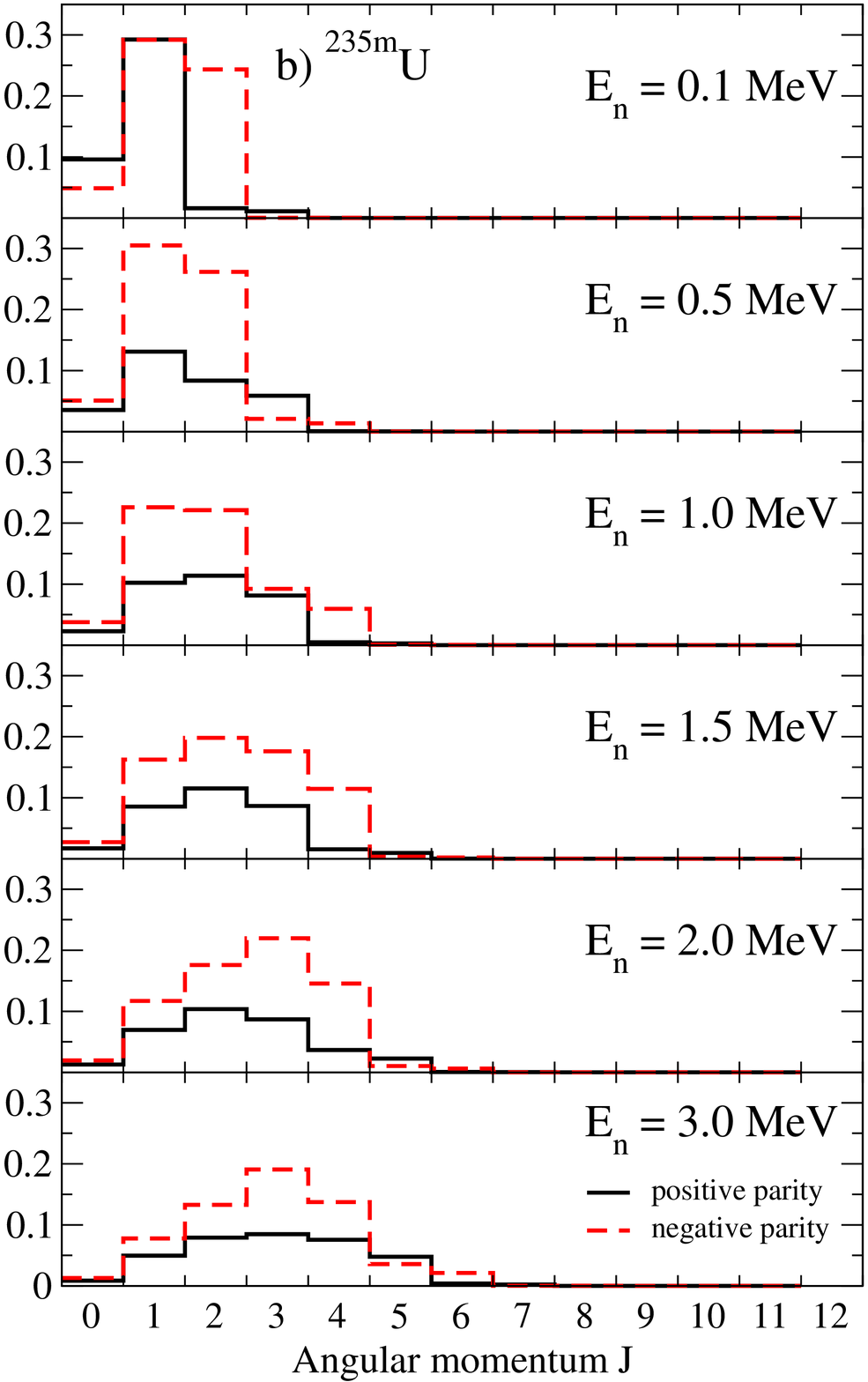}
}
\caption[Spin-parity distributions for the compound nucleus $^{236}$U]
{(Color online) \jpi distributions for the compound nucleus $^{236}$U, following its production in the a) $n$ +$^{235}$U and b) $n$ +$^{235m}$U reactions, for various neutron energies.  Shown are the probabilities for occupying compound-nuclear states with specific spin and parity in neutron-induced reactions, as given in the second line of Eq.~\ref{Eq:Lucky}.  The calculations were carried out using the Flap 2.2 optical model (see appendix). 
}
\label{fig:jpi_n235U}
\end{figure*}

While the strong dependence of the $\gamma$-ray yields on the \jpi distribution of the compound nucleus makes the extraction of a $(n,\gamma)$ cross section from a surrogate experiment difficult, this sensitivity also provides an opportunity for obtaining information on the spin-parity distribution of the decaying nucleus from an observation of the associated $\gamma$ rays.  Measurements of yields for various individual $\gamma$ rays will provide stringent tests for theoretical predictions of the formation and decay of a compound nucleus produced in a surrogate reaction.


\section{Method of the Study}
\label{sec_method}

We designed several simulations to test the surrogate method in the \we limit and two variants of the ratio method.  
We selected rare-earth and actinide isotopes for which direct cross-section measurements are available to compare against: $^{155,157}$Gd~\cite{X4_155GdngCS} and $^{233,235}$U~\cite{Hopkins:62}.  For each nucleus, we carried out a full Hauser-Feshbach calculation of the neutron-induced reaction and calibrated the model parameters to give an overall good fit of the known neutron resonance spacings, average radiative widths, and \nga cross sections; for the uranium nuclei, the fits included the (n,f) cross sections.  
The quality of the fits is very good, as can be seen in Fig.~\ref{fig:U235RefXS} for \ufive (the \uthree cross sections are of similar quality) and in Fig.~\ref{fig:UngSimulWE_Gd} for $^{155,157}$Gd.  All calculations were carried out with a modified version of the Hauser-Feshbach code {\sc Stapre}~\cite{Uhl:76,Strohmaier:80}  
that allowed us to extract the branching ratios for capture as a function of spin and parity of the initially formed compound nucleus, $G_{\gamma}^{CN}(E,J,\pi)$.  

\begin{figure}[htb]
\centering
\resizebox{1.0\columnwidth}{!}{
    \includegraphics[viewport=80 12 590 730,clip,angle=270]{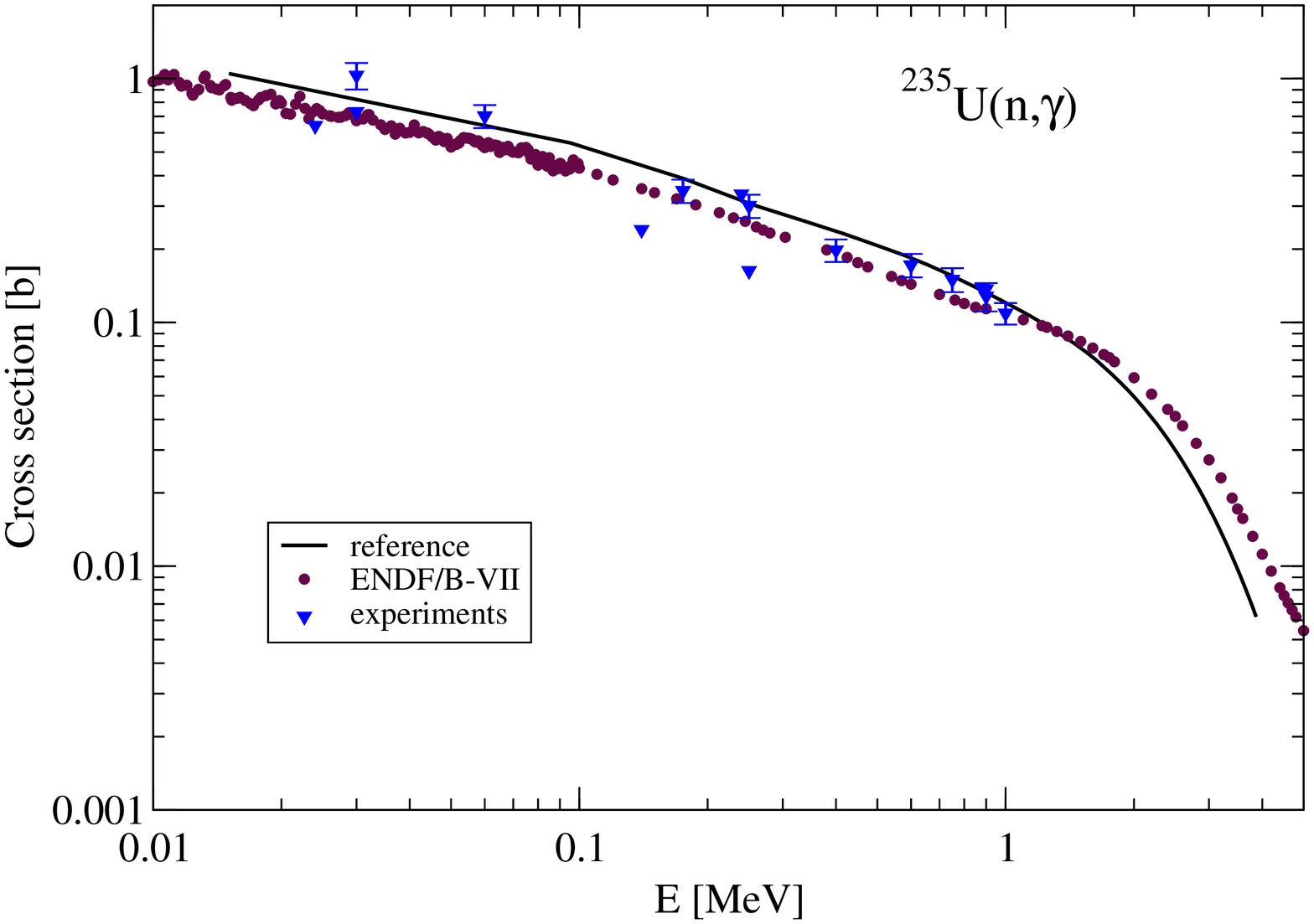}
}
\hfill
\resizebox{1.0\columnwidth}{!}{
    \includegraphics[viewport=80 12 580 730,clip,angle=270]{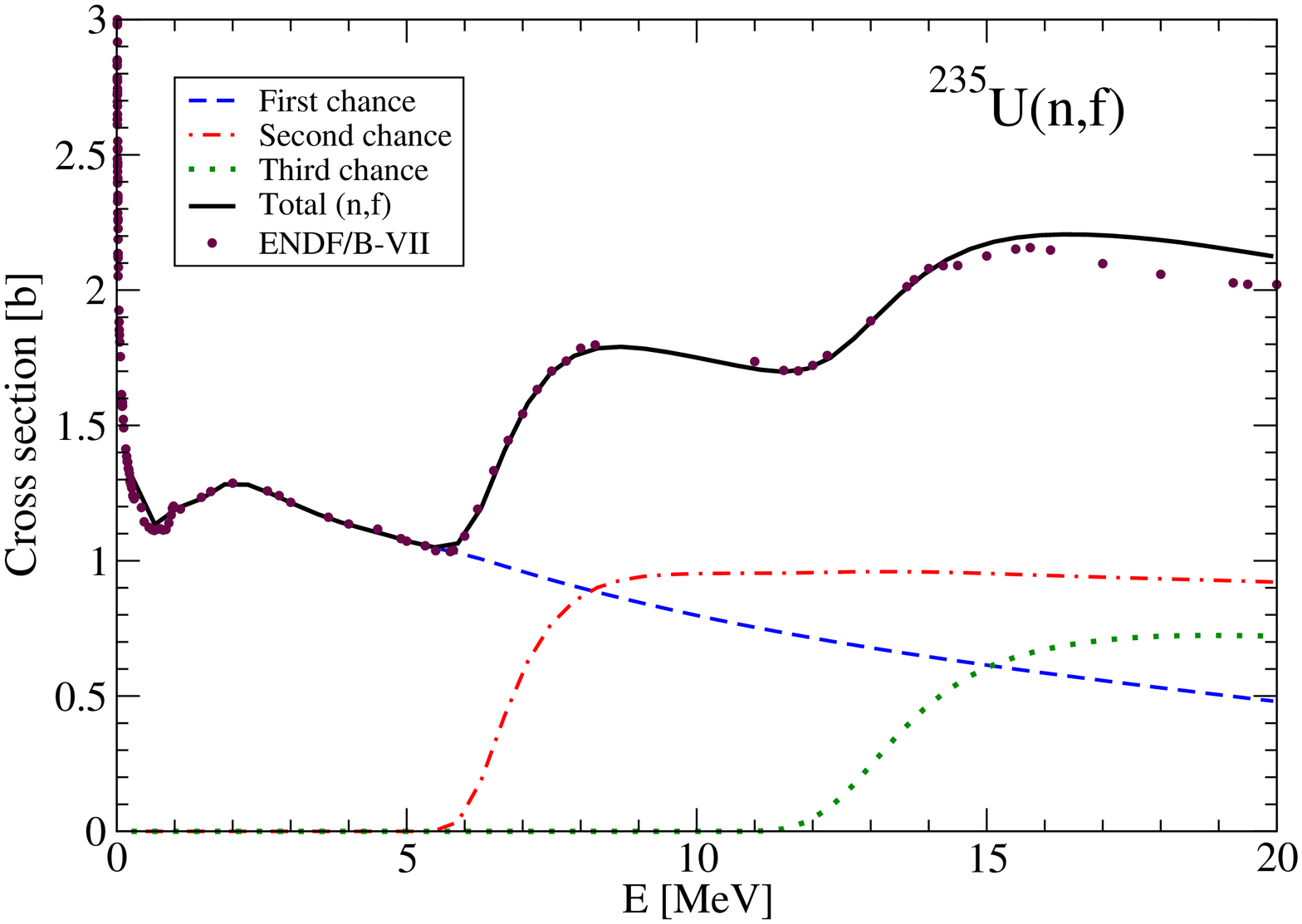}   
}
\caption[Fits to the \ufivex(n,f) and \ufivex\nga cross sections]{(Color online) Fits to the \ufivex(n,f) and \ufivex\nga cross sections, as function of neutron energy.  We started from the (n,f) calculations discussed in Ref.~\cite{EscherDietrich:06PRC}, and slightly adjusted the model parameters to better reproduce the available \nga data and {\sc ENDF/B-VII} evaluation for the \uthreex\nga and \ufivex\nga cross sections.  The filled circles correspond to the ENDF/B-VII evaluation~\cite{ENDFb7:06}.  Little data exists for the \ufivex\nga case; three such data sets, Refs.~\cite{Hopkins:62,Spivak:56,Andreev:58}, are plotted in the upper panel.
}
\label{fig:U235RefXS}
\end{figure}

Employing the \we assumption in the analysis of surrogate reactions for which this approximation is not valid will result in extracted cross sections that deviate from the desired true cross section.  
The effect of the spin-parity mismatch between the desired and surrogate reactions on the cross section extracted from a Weisskopf-Ewing analysis can be simulated by employing the calculated $G^{CN}_{\gamma}(E,J,\pi)$ and a range of possible surrogate spin-parity distributions.  We consider several schematic, energy-independent distributions $F^{CN (p)}_{\delta}(E,J,\pi)$, where the $p$ labels the distribution under consideration, and calculate simulated surrogate coincidence probabilities 
\bea
P^{(p)}_{\delta,\gamma}(E) &=& \sum_{J,\pi} F_{\delta}^{CN (p)}(E,J,\pi) G_{\gamma}^{CN}(E,J,\pi) \; .
\label{eq:simRatio}
\eea
Treating the latter like an experimental result, one can calculate $\sigma^{WE}_{(n,\gamma)}(E)$ = $\sigma^{CN}_{n+target}(E)$ $P_{\delta,\gamma}(E)$, which corresponds to a surrogate analysis in the \we approximation (cf.\ Eq.~\ref{Eq:WElimitXSec}).
The compound-nucleus formation cross section is $\sigma^{CN}_{n+target}(E) =  \sum_{J\pi} \sigma^{CN}_{n+target} (E,J,\pi)$, where the individual $\sigma^{CN}_{\alpha} (E,J,\pi)$ were taken to be those used for the fits to the direct cross-section measurements.
The resulting simulated cross sections can then be compared to the reference cross sections calculated from the full Hauser-Feshbach theory.
The validity of the ratio approaches can be studied by the same technique.


\section{Results for the actinide region}
\label{sec_actinides}

Given the spin-parity dependence of the $\gamma$-branching ratios $G_{\gamma}^{CN}(E,J,\pi)$ shown in Fig.~\ref{fig:gamProbsU236}, we expect a \we analysis of a surrogate experiment to be of limited value for obtaining an \nga cross section, unless the surrogate spin-parity distribution is similar to the distribution produced in the desired reaction.
Here, we study the accuracy that can be achieved with the \we approximation when a spin-parity mismatch exists between the desired and surrogate reactions. We also investigate whether the accuracy can be improved by using the surrogate ratio method.

\subsection{Validity of the Weisskopf-Ewing approximation for (n,$\gamma$) on actinide targets}
\label{sec_actinides_WE}

To simulate a range of surrogate reactions, we consider four schematic, energy-independent distributions $F^{CN (p)}_{\delta}(E,J,\pi)$.  The first three are distributions A, B, D employed in our study of the fission channel, as discussed in Ref.~\cite{EscherDietrich:06PRC}.  They are shown in Fig.~\ref{fig_jdistrib_uran}b.  Distribution C of that study is not considered here, as the reaction mechanisms employed in most recent surrogate experiments are not expected to populate such high angular-momentum states.  Instead, we have added distribution ABB, which we extracted from a (d,p) prediction made by Andersen, Back, and Bang~\cite{Andersen:70}.  This distribution is shown in Fig.~\ref{fig_jdistrib_uran}a.  In ABB, the $J$ distributions are parity dependent; in D, A, and B they are not.  We calculate simulated surrogate coincidence probabilities $P^{(p)}_{\delta\gamma}(E)=$ $\sum_{J,\pi}F^{CN (p)}_{\delta}(E,J,\pi)G^{CN}_{\gamma}(E,J,\pi)$ for the four different distributions ($p = ABB,D,A,B$) and obtain -- in the Weisskopf-Ewing approximation -- the $^{235}$U(n,$\gamma$) cross sections indicated in Fig.~\ref{fig:UngSimulWE}a.  Analogously, one obtains the $^{233}$U(n,$\gamma$) cross sections shown in panel b) of the figure.  In both cases the compound-nuclear formation cross section of Fig.~\ref{fig_sigreac_fsd} (see appendix) was used.
These calculations are compared with the ``reference cross section", which was obtained from the fit discussed in Section~\ref{sec_method} and is shown in Fig.~\ref{fig:U235RefXS}.

\begin{figure}[htb]
\vspace{0.6cm}
\centering
\resizebox{1.0\columnwidth}{!}{
    \includegraphics[angle=270,viewport=80 15 530 730,clip]{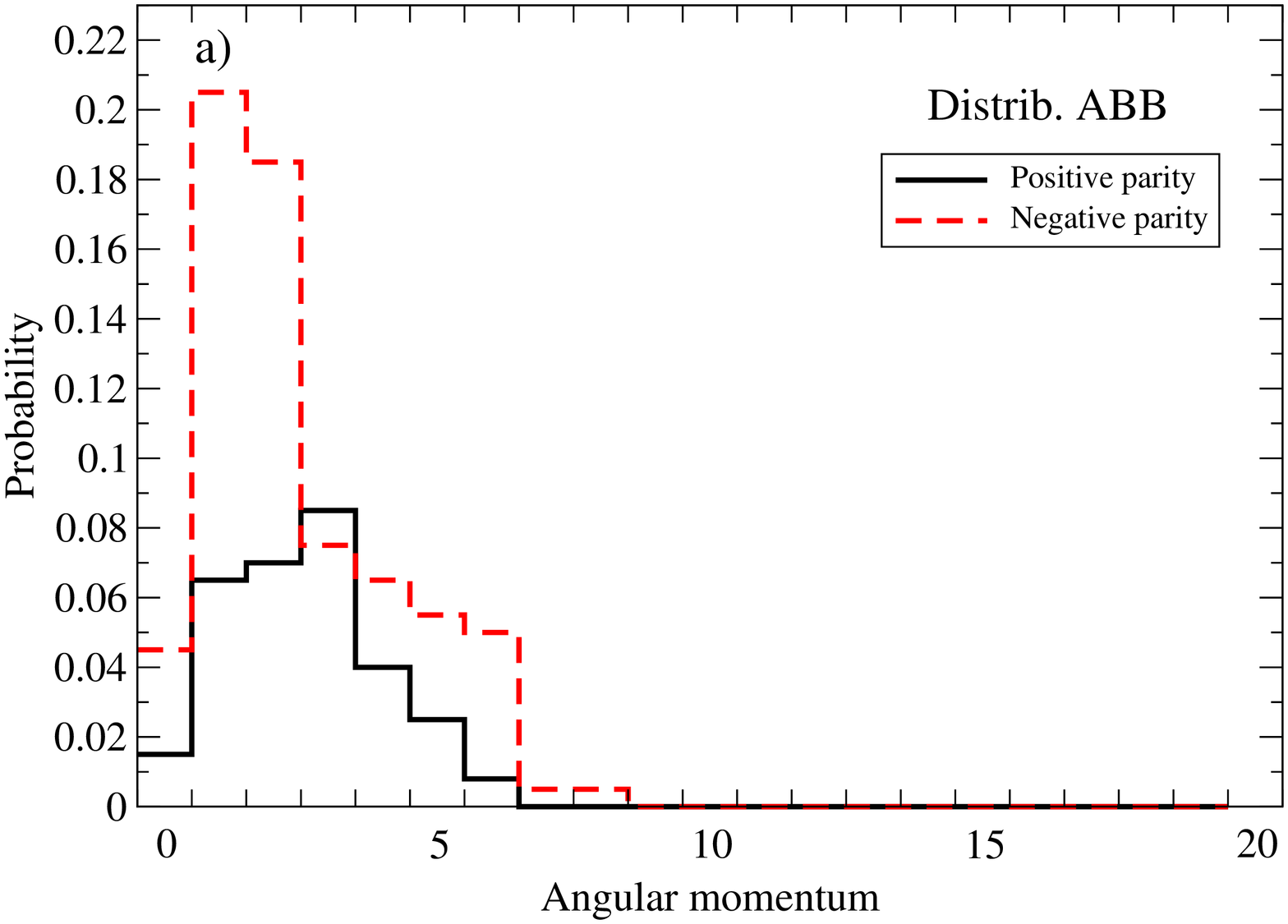}
}
\hfill
\resizebox{1.0\columnwidth}{!}{
    \includegraphics[angle=270,viewport=80 15 590 730,clip]{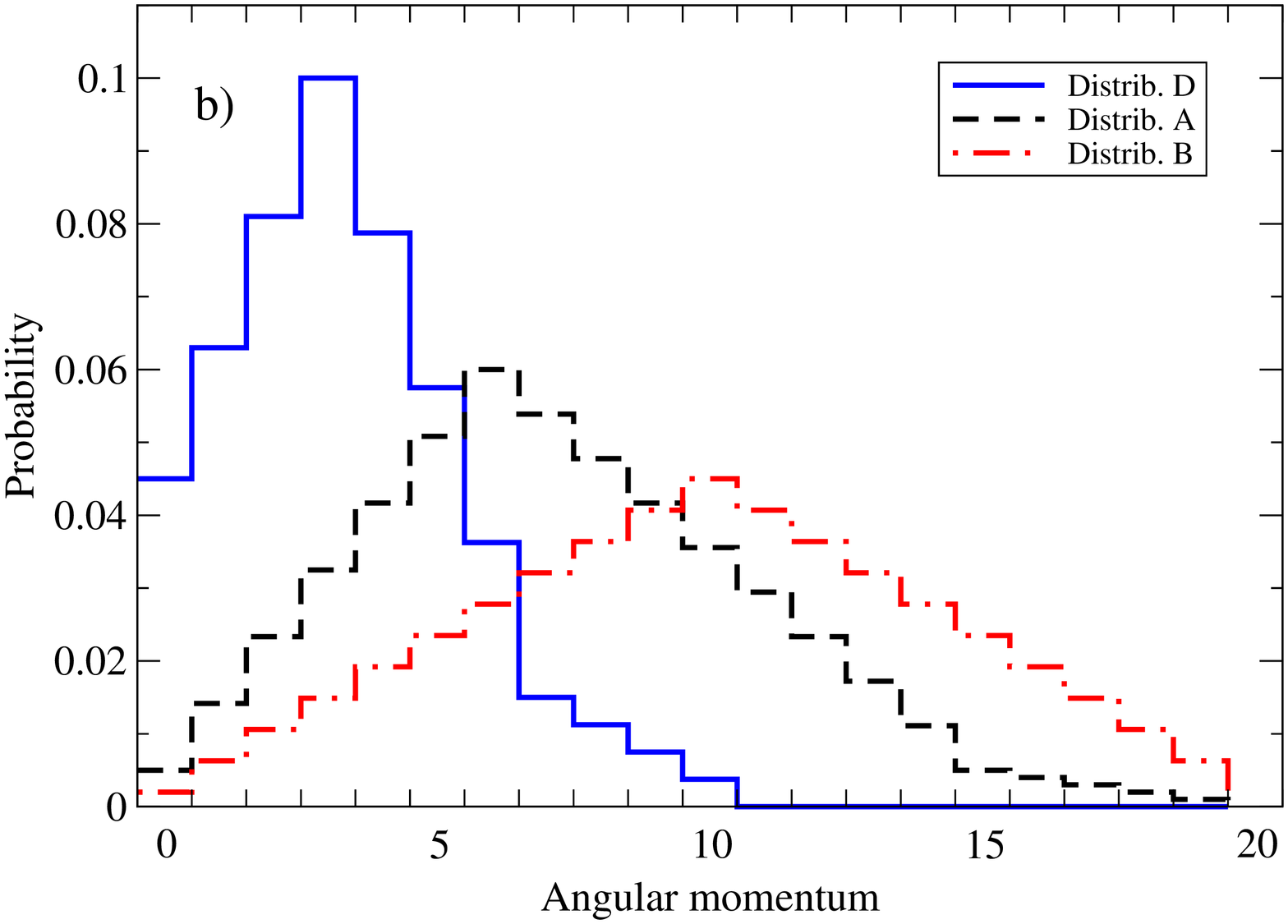}
}
\caption[Schematic spin-parity distribution distributions a) ABB and b) D, A, B.]{
(Color online) Schematic spin-parity distributions a) ABB and b) D, A, B.}
\label{fig_jdistrib_uran}
\vspace{0.1cm}
\end{figure}

We observe that for energies above $E_n \approx$ 0.6 MeV the energy dependence of the radiative capture cross section is reasonably well reproduced by the Weisskopf-Ewing simulation, while the absolute magnitudes are strongly dependent on the assumed spin-parity distribution in the surrogate reaction.  Distributions ABB and D lead to results that are very close to the \ufivex(n,f) reference cross section, while the cross sections associated with distributions A and B are too large by about 40\% and 200\%, respectively.  Distributions ABB and D also yield very good agreement with the  \uthreex(n,f) cross section, but the cross sections extracted for distributions A and B are too large by roughly 20\% and 50\%, respectively.

Since the deviations between the extracted and reference cross sections are different for the \uthree and \ufive examples considered here, it is unlikely that a simple (rescaling) procedure can be identified that corrects for the spin-parity mismatch, which is neglected in the \we approximation.

\begin{figure}[htb]
\vspace{0.5cm}
\centering
\resizebox{1.0\columnwidth}{!}{
    \includegraphics[angle=270,viewport=80 15 530 730,clip]{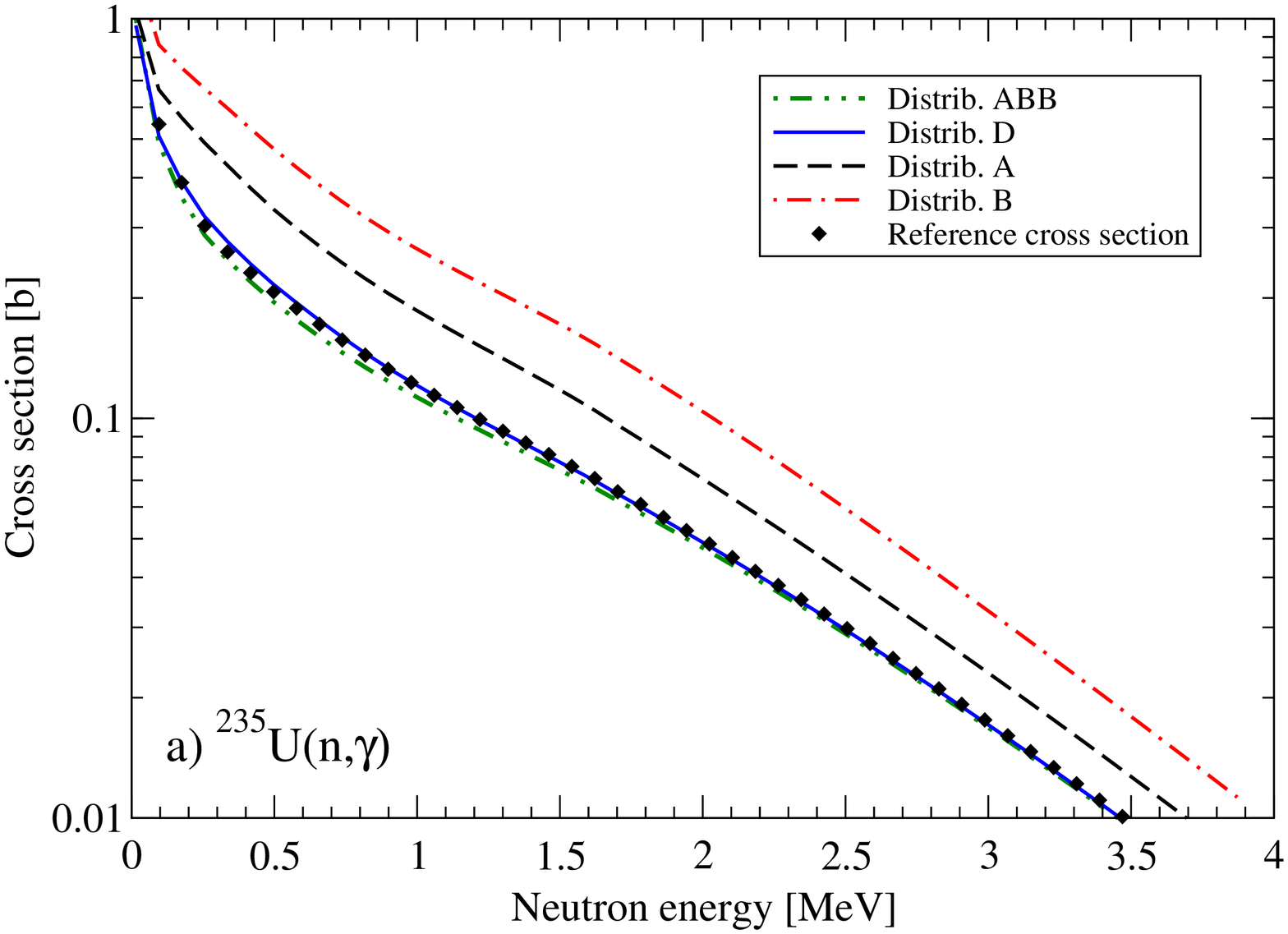}
}
\hfill
\resizebox{1.0\columnwidth}{!}{
    \includegraphics[angle=270,viewport=80 15 590 730,clip]{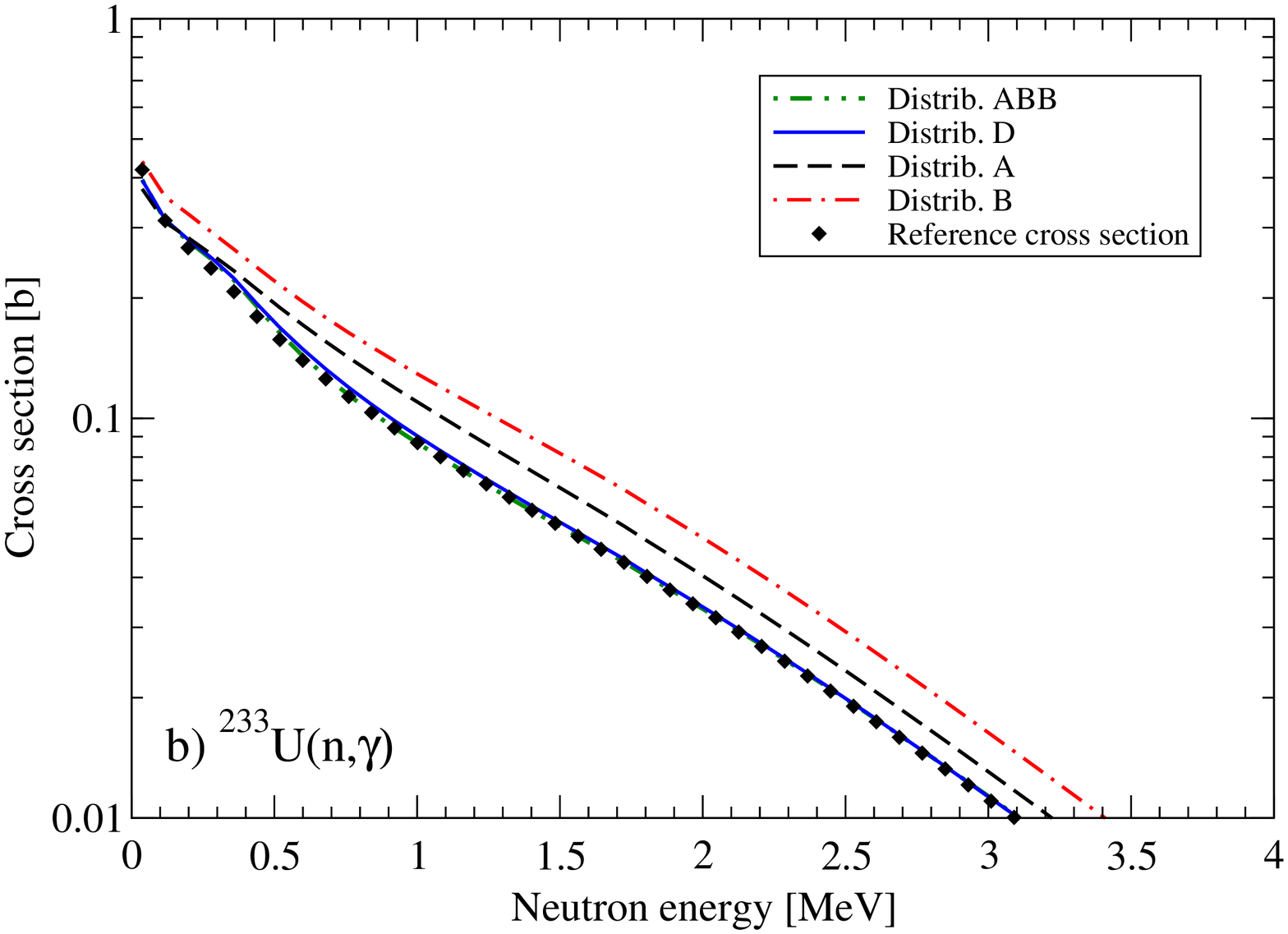}
}
\caption[Weisskopf-Ewing estimates for the a) \ufivex(n,$\gamma$) and b) \uthreex(n,$\gamma$) cross sections from simulated surrogate experiments]
{(Color online) Weisskopf-Ewing estimates for the a) \ufivex(n,$\gamma$) and b) \uthreex(n,$\gamma$) cross sections, extracted from analyses of simulated surrogate experiments, for four different compound-nuclear \jpi distributions. For comparison, the reference cross section, which was obtained by adjusting the parameters for the Hauser-Feshbach calculation to reproduce direct measurements or cross section evaluations is shown as well.
}
\label{fig:UngSimulWE}
\end{figure}

The calculated $\gamma$-decay probabilities $G^{CN}_{\gamma}(E,J,\pi)$ shown in Fig.~\ref{fig:gamProbsU236} help us understand discrepancies between the reference cross section and those extracted from a Weisskopf-Ewing analysis of the surrogate data:  If the surrogate reaction populates the relevant compound nucleus, e.g.\  $^{236}$U, with a spin-parity distribution that contains larger angular-momentum values than the population relevant to the neutron-induced reaction, then the measured decay probability $P_{\delta\gamma}(E)$ of Eq.~\ref{Eq:SurReact} contains larger contributions from those $G^{CN}_{\gamma}(E,J,\pi)$ associated with large $J$ values than the cross section expression for the desired $(n,\gamma)$ reaction does. Consequently, the cross section extracted by using the Weisskopf-Ewing assumption and approximating $P_{\delta\gamma}(E) \approx G^{CN}_{\gamma}(E)$, gives too large a result.  The same will hold true for other surrogate mechanisms that produce the compound nucleus with spin-parity distributions that are shifted to larger $J$ values relative to the distribution found in the neutron-induced reaction.
We note that distribution ABB arises from a theoretical calculation for a specific type of direct reaction, while the others are purely schematic and designed specifically for the kind of sensitivity study presented here.

We conclude that the Weiskopf-Ewing approximation does not lead to a satisfactory estimate of the radiative capture cross section unless the spin-parity distribution in the compound nucleus is adequately known and a surrogate reaction mechanism and experimental conditions can be identified and devised that approximately reproduce the spin-parity distribution of the desired reaction.

\subsection{Validity of the ratio approximations for (n,$\gamma$) reactions on actinide targets}
\label{sec_actinides_ratios}

One may try to reduce the uncertainties seen in the cross sections obtained using the \we analysis by employing the surrogate ratio method.
We treat the \ufivex\nga cross section as the `unknown' cross section to be determined from a surrogate ratio analysis. In applications of the ratio method, the unknown cross section is determined relative to a suitable known cross section.  Here we select \uthreex\nga as the `known' cross section for the purpose of testing the external surrogate ratio method.  For probing the internal surrogate ratio method, we use the \ufivex\nf cross section as the known quantity.

\paragraph{External Surrogate Ratio for \nga cross sections.}
To test the external ratio (ESR) method, we determine the \ufivex\nga cross section from a ratio analysis of the simulated surrogate data for the decays of \usix and \ufour by $\gamma$ emission.
The `known' cross section is taken to be the \uthreex\nga reference cross section.
The results are shown in Fig.~\ref{fig:UngSimulExSR}a.  In order to better display the differences for the selected spin-parity distributions, we also show the cross section ratios $\sigma$[\ufivex\ngax] / $\sigma$[\uthreex\ngax] obtained for the four schematic \jpi distributions and the reference cross sections (see Fig.~\ref{fig:UngSimulExSR}b).
The dependence on the spin-parity distribution is reduced relative to the results for the \we approximation but is still quite large for the distributions (A, B) having a very large high-spin component.
In particular, the shape of the reference cross section is approximately reproduced for $E_n >$ 1 MeV, but the magnitudes of the results extracted from the external ratio analysis are too large by roughly 20\% (for distribution A) to 40\% (for distribution B). 
The disagreement between the cross sections extracted using the ESR analysis and the reference result decreases with increasing energy, but even at $E_n \approx 3-4$ MeV, the \we limit does not seem to apply.  This is different from what has been found in simulations of \nga cross sections for zirconium~\cite{Forssen:07}. (For that case, the \we limit was reached around 2.5-3 MeV.)
For lower energies, $E_n <$ 0.5 MeV, the discrepancies increase, even the shape of the cross section is no longer properly reproduced.
The results for distributions ABB and D are within 5\% of the reference ratio, {\it i.e.\ } for surrogate spin-parity distributions that are similar to those found in the desired, neutron-induced reaction, the external surrogate ratio approach gives results that are in good agreement with the expected cross section.

\begin{figure}[htb]
\vspace{0.5cm}
\centering
\resizebox{1.0\columnwidth}{!}{
    \includegraphics[angle=270,viewport=80 15 530 730,clip]{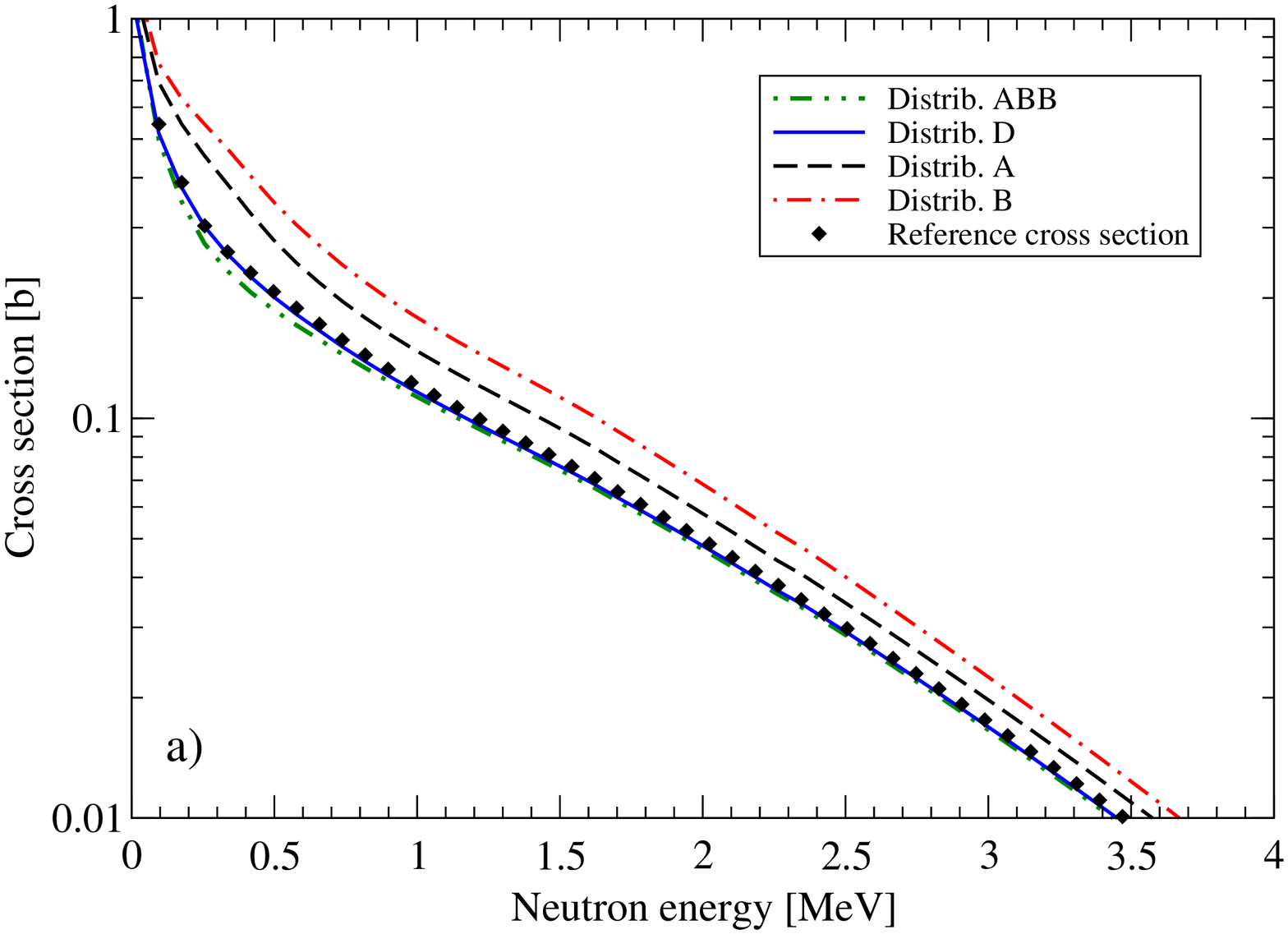}
}
\resizebox{1.0\columnwidth}{!}{
    \includegraphics[angle=270,viewport=80 15 590 730,clip]{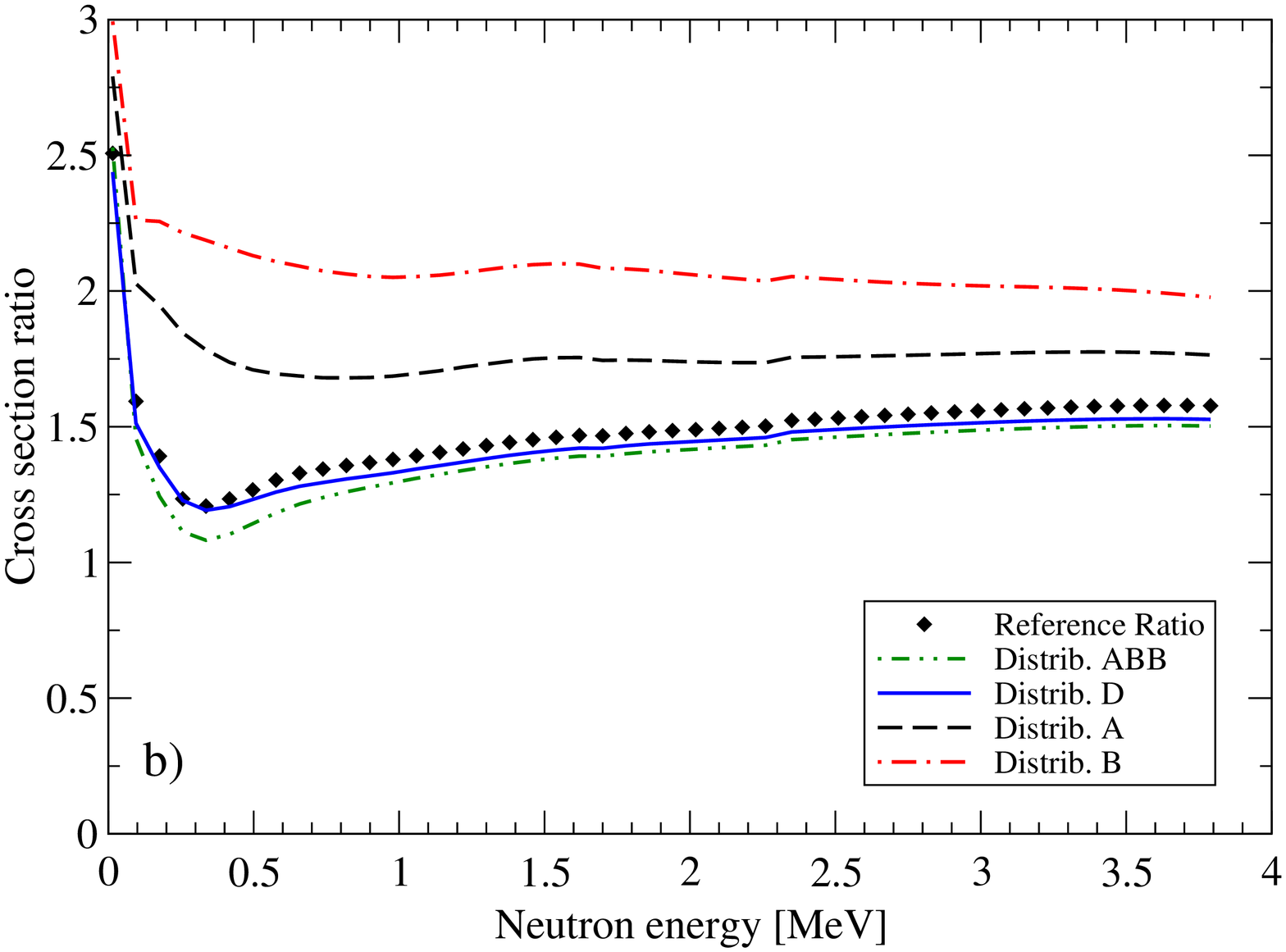}
}
\caption[External surrogate ratio estimates for \ufivex\ngax]{(Color online) External surrogate ratio estimates for the \ufivex\nga cross section, extracted from analyses of simulated surrogate experiments, compared to the reference cross section.  Panel a) shows the cross section result of a simulated external surrogate ratio analysis, while panel b) shows the ratio of the cross sections, $\sigma$[\ufivex\ngax] / $\sigma$[\uthreex\ngax]. Four different compound-nuclear \jpi distributions were considered.
}
\label{fig:UngSimulExSR}
\end{figure}

\paragraph{Internal Surrogate Ratio for \nga cross sections.}
To test the internal ratio (ISR) method, we determine the \ufivex\nga cross section from a ratio analysis of the simulated surrogate data for the decays of \usix via fission and $\gamma$ emission; the `known' cross section is taken to be the \ufivex\nf reference cross section.
The results obtained are shown in Fig.~\ref{fig:UngSimulInSR}.  Panel a) gives the \ufivex\nga cross sections extracted from the ISR analysis of simulated surrogate experiments, compared to the reference cross section. 
For $E_n > 0.5$ MeV, the pattern is similar to those found for the (absolute) \we and ESR analyses, with some improvement over the \we result and a spin-parity dependence of the cross section that is similar to that seen for the ESR case.  Panel b), which shows the ratio of the \ufivex\nga cross section to the \ufivex\nf cross section, illustrates the drop in the cross-section ratios with increasing energy.  Results for distributions D and ABB are nearly indistinguishable from each other in both panels of the figure.
Cross sections associated with these distributions differ from the reference cross section by only about 3-10\%

\begin{figure}[htb]
\vspace{0.5cm}
\centering
\resizebox{1.0\columnwidth}{!}{
    \includegraphics[angle=270,viewport=80 15 530 730,clip]{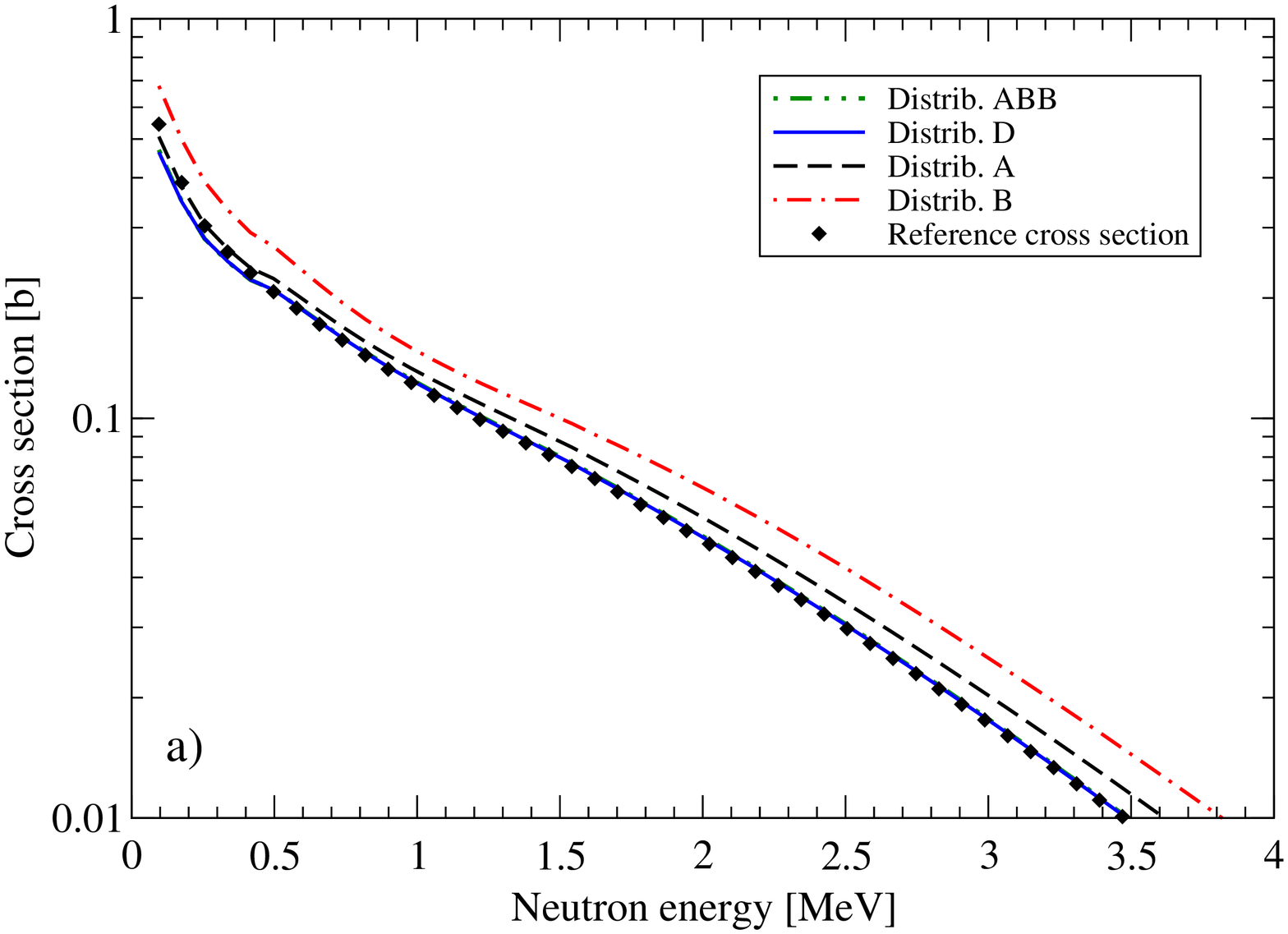}
}
\resizebox{1.0\columnwidth}{!}{
    \includegraphics[angle=270,viewport=80 15 590 730,clip]{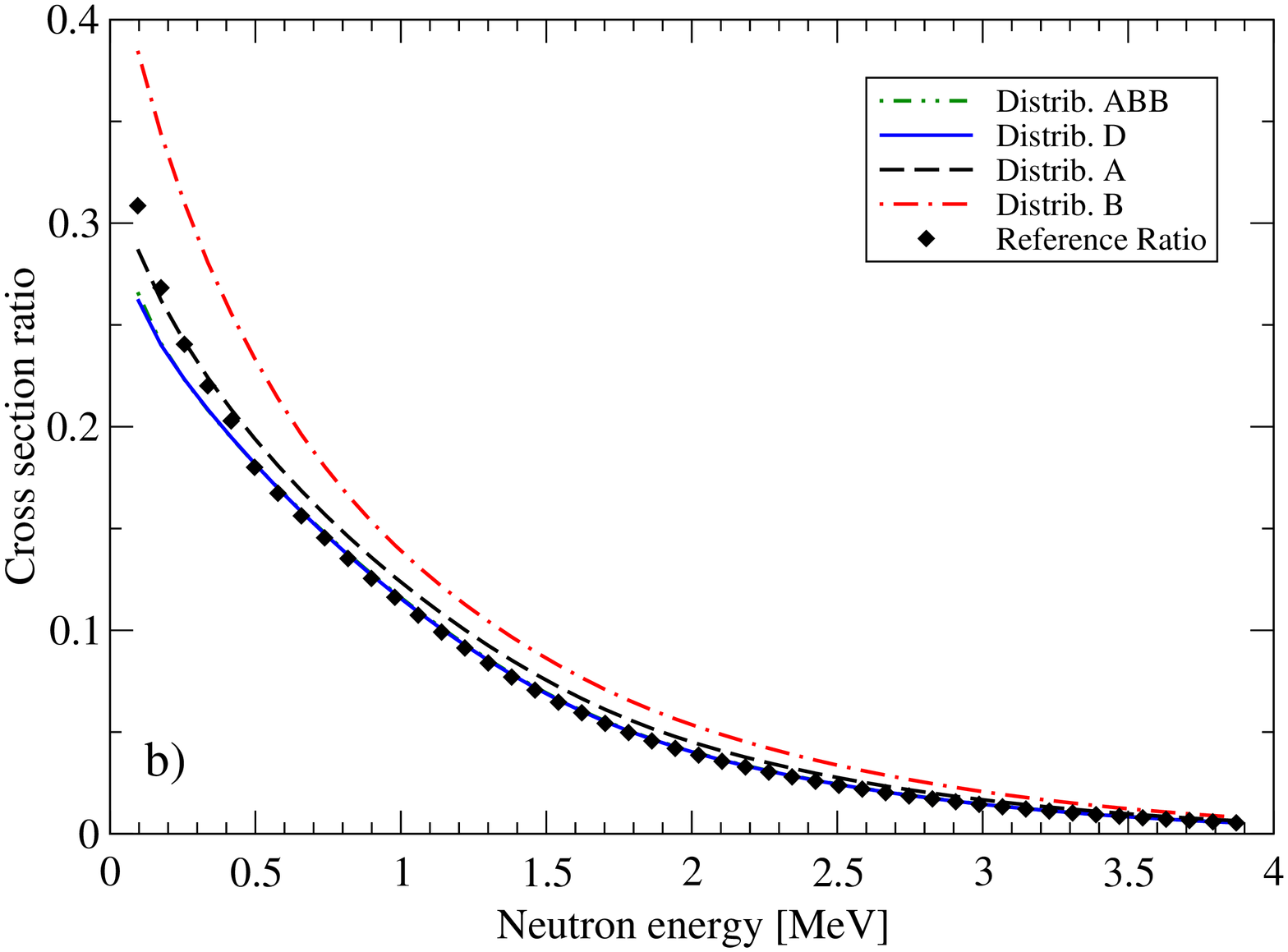}
}
\caption[Internal surrogate ratio estimates for \uthreex(n,$\gamma$)]{(Color online) Internal surrogate ratio estimates for the \ufivex\nga cross section, extracted from analyses of simulated surrogate experiments, compared to the reference cross section.   Panel a) shows the cross section result, while panel b) shows the ratio of the cross sections, $\sigma$[\ufivex\ngax] / $\sigma$[\ufivex(n,f)].
}
\label{fig:UngSimulInSR}
\end{figure}


\section{Results for the rare-earth region}
\label{sec_rareearths}

A study of the rare-earth isotopes is valuable, as several recent surrogate experiments have focused on this region~\cite{Scielzo:09ip,Hatarik:07cnr,Hatarik:10,Goldblum:08,Bleuel_xxxx}.
The level densities typically encountered in this mass region are much higher than those found near closed-shells.  Consequently, we expect the $\gamma$-branching ratios for this region to be less sensitive to spin effects than the ratios found in the recent study of the zirconium isotopes~\cite{Forssen:07}.
Moreover, it becomes possible to study features generally relevant to \nga applications without the added complication of the competing fission channel.

We selected the  $^{155,157}$Gd\nga cross sections, which have been measured directly, for neutron energies up to about 2.5 MeV (Fig.~\ref{fig:UngSimulWE_Gd}).  The target nuclei,  $^{155,157}$Gd, have ground-state spin and parity \jpi=$3/2^+$, and similar deformations.  The compound nuclei of interest,  $^{156,158}$Gd, are both even-even, have similar deformations, level structure, and decay schemes.  Various stable isotopes exist near the compound nuclei of interest, which makes experimental studies of this region feasible: surrogate experiments employing $(^3$He,$\alpha$)~\cite{Bleuel_xxxx} and inelastic scattering reactions~\cite{Scielzo:09ip} to produce compound  $^{156,158}$Gd nuclei have recently been carried out.

\begin{figure*}[htb]
\centering
\resizebox{1.0\columnwidth}{!}{
    \includegraphics[viewport=80 15 520 730,clip,angle=270]{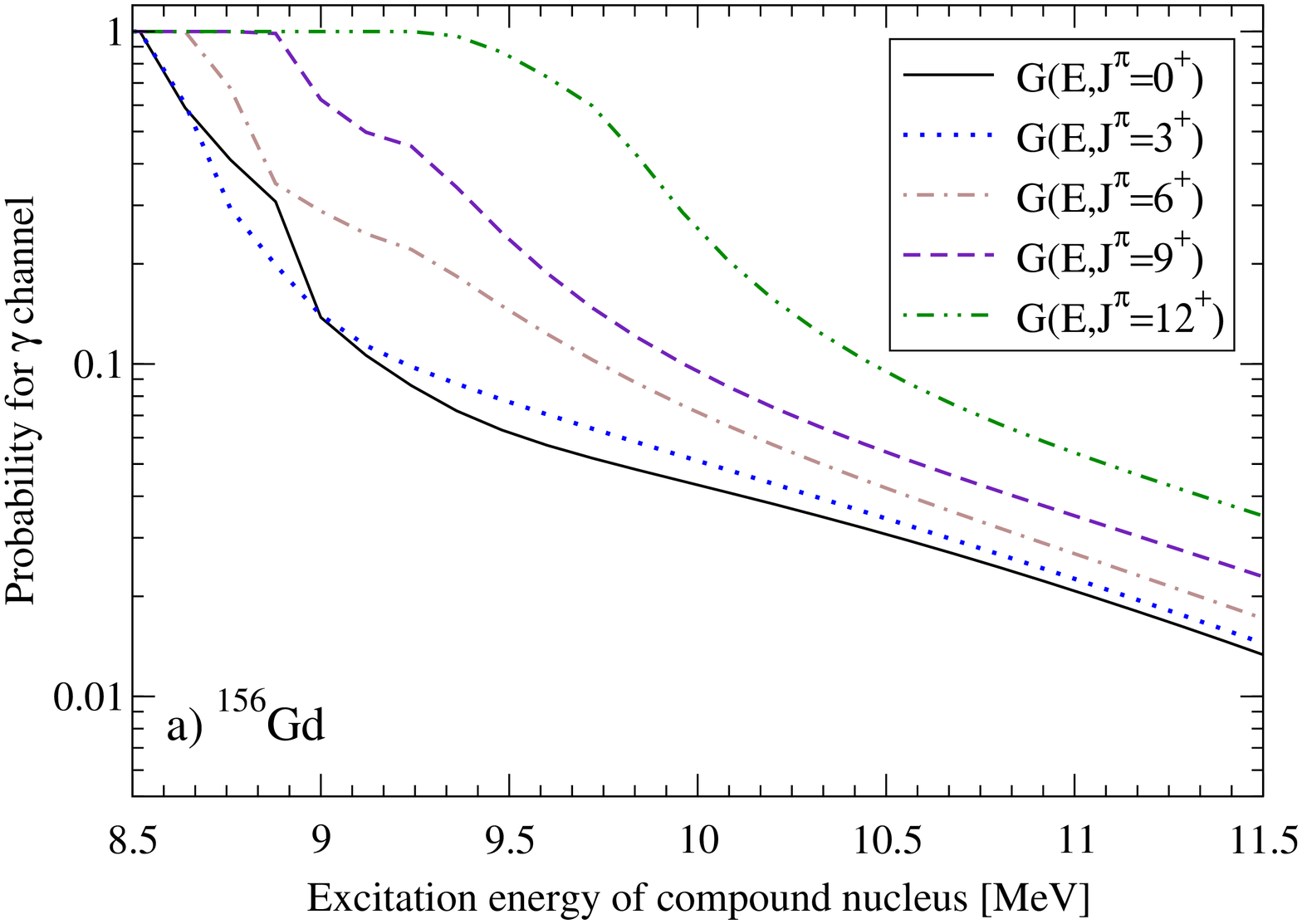}
}
\resizebox{1.0\columnwidth}{!}{
    \includegraphics[viewport=80 15 520 730,clip,angle=270]{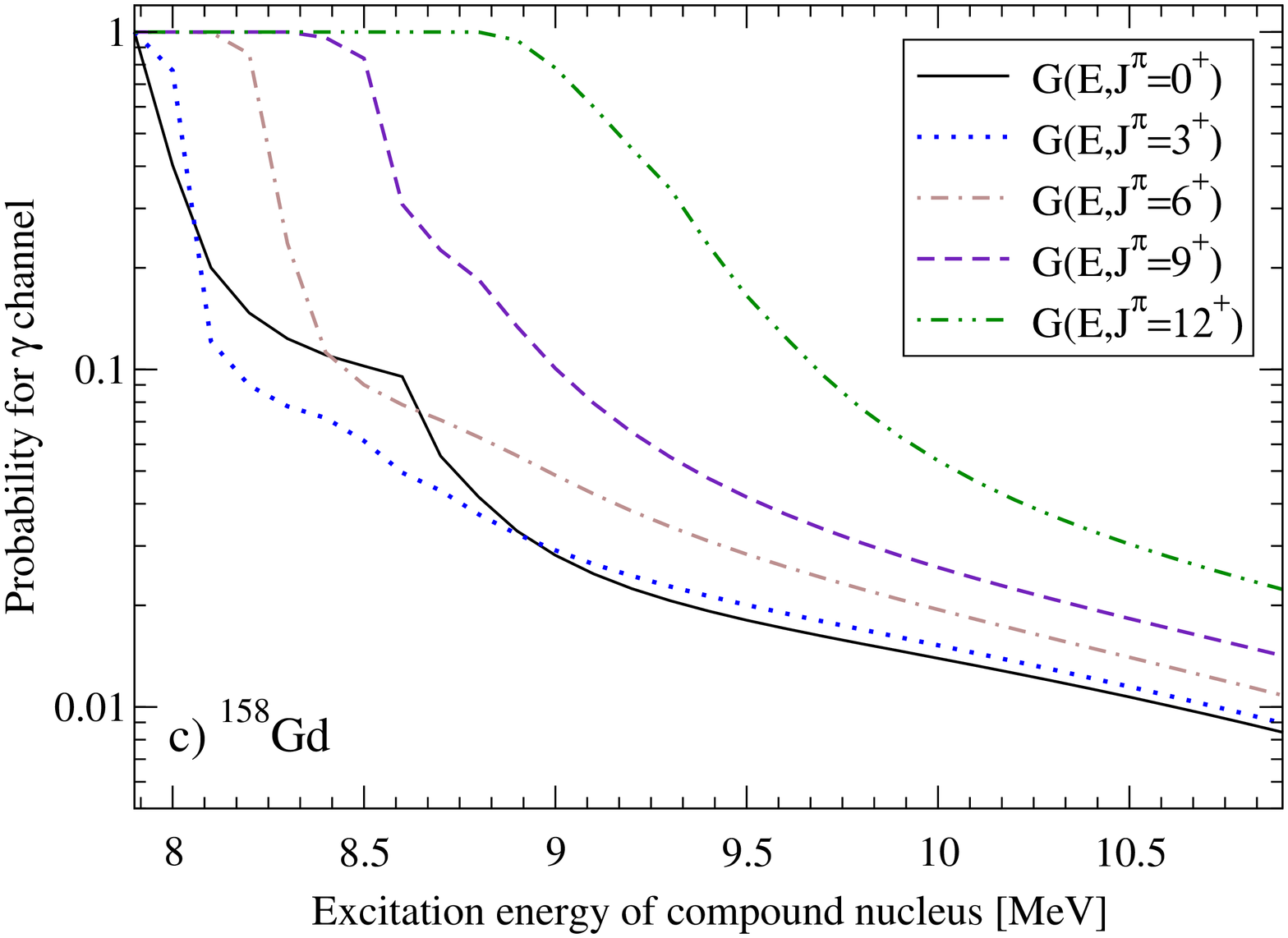}
}
\hfill
\resizebox{1.0\columnwidth}{!}{
    \includegraphics[viewport=80 15 610 730,clip,angle=270]{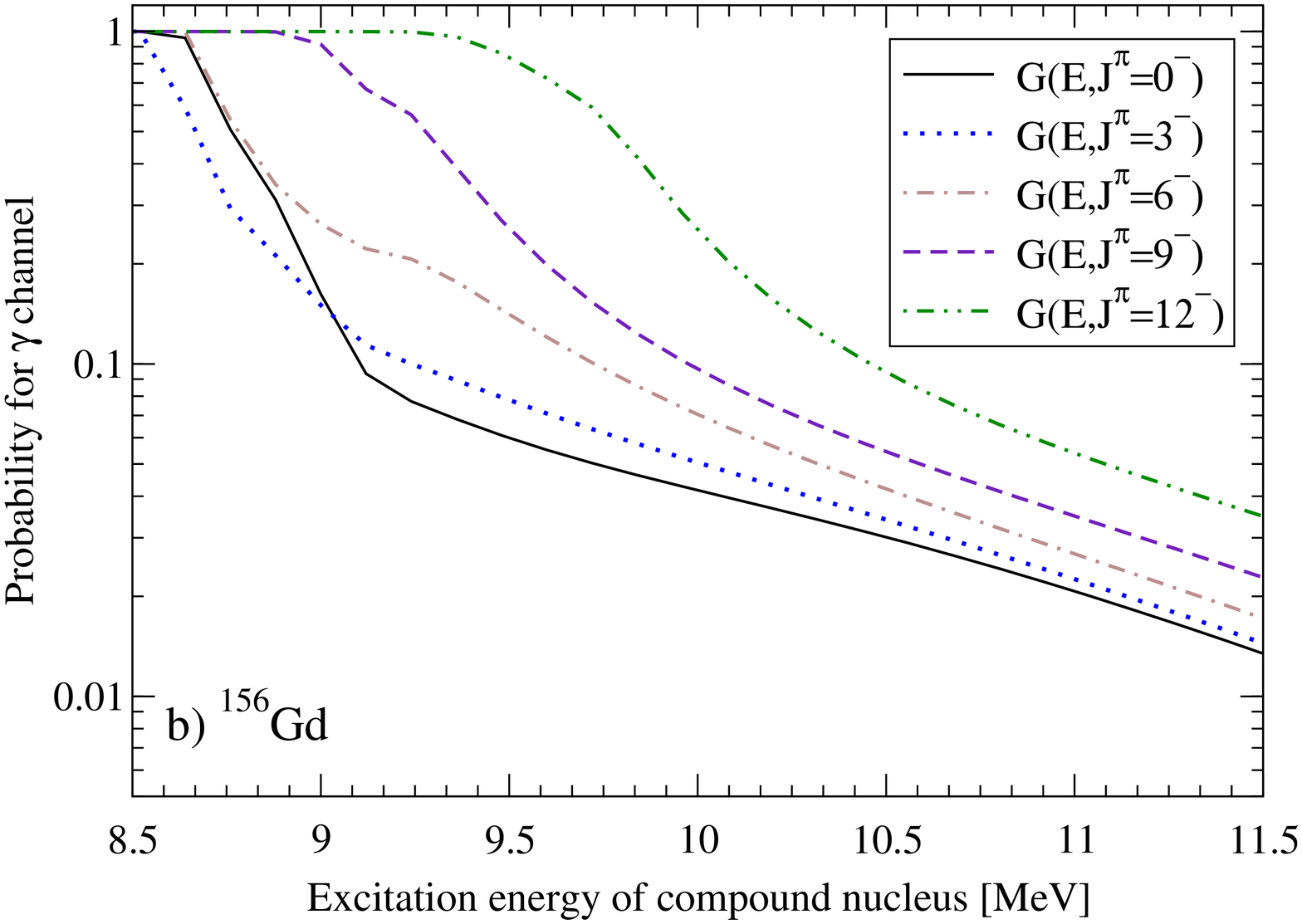}  
}
\resizebox{1.0\columnwidth}{!}{
    \includegraphics[viewport=80 15 610 730,clip,angle=270]{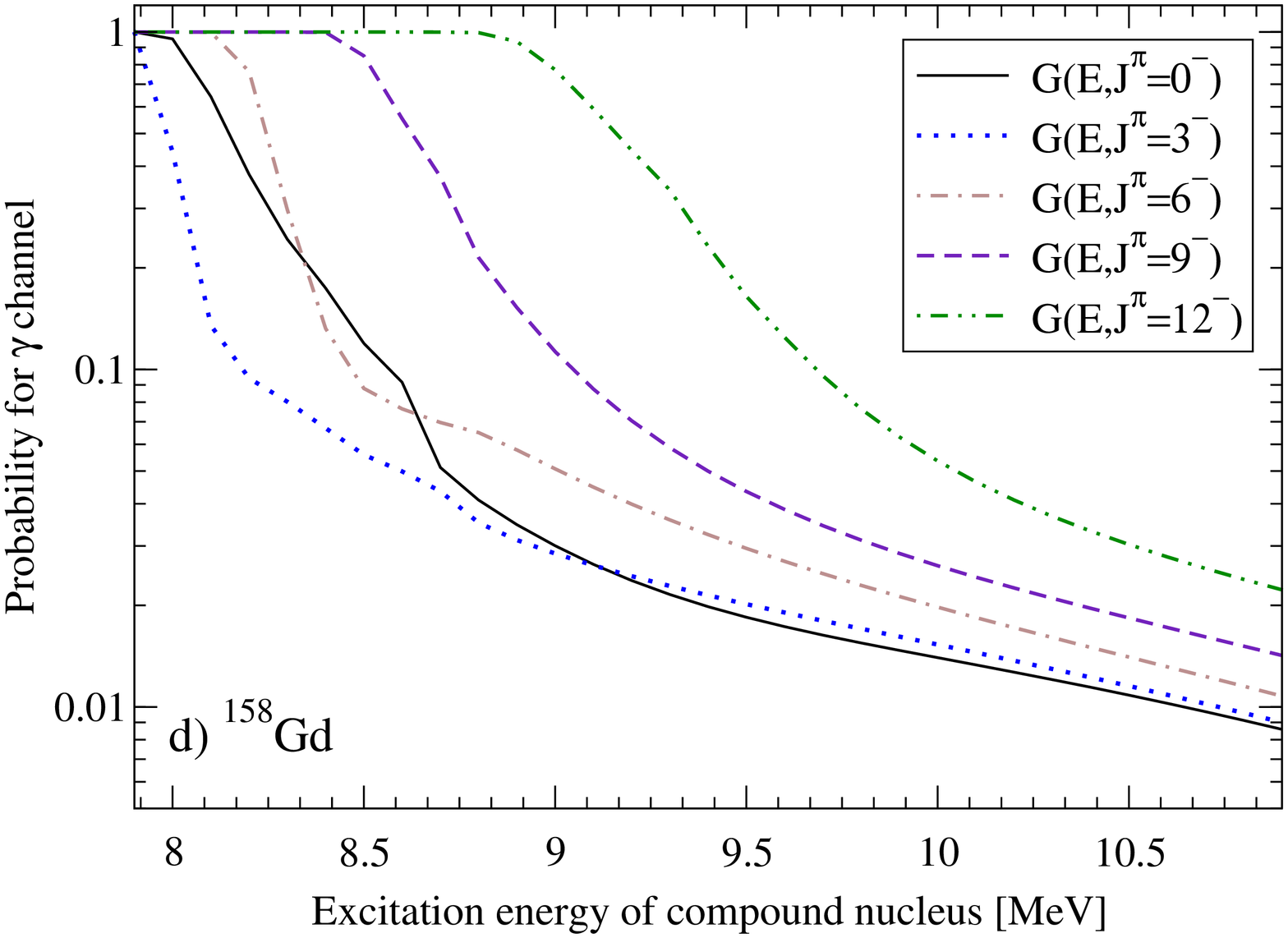}  
}
\caption[Calculated $\gamma$-decay probabilities $G^{CN}_{\gamma}(E,J,\pi)$, for Gd]{(Color online) Calculated $\gamma$-decay probabilities $G^{CN}_{\gamma}(E,J,\pi)$, for $^{156,158}$Gd. Shown is the probability that the compound nucleus, when produced with a specific \jpi combination, decays via the $\gamma$ channel.  Positive-parity decay probabilities are shown in panels a) and c), and negative-parity decay probabilities are shown in panels b) and d).
}
\label{fig:gamProbsGd}
\end{figure*}

In Fig.~\ref{fig:gamProbsGd}, we show $\gamma$ branching ratios $G^{CN}_{\gamma}(E,J,\pi)$ for the decay of \gdsix and $^{158}$Gd, for excitation energies corresponding to neutrons with $E_n = 0-3$ MeV ($S_n$(\gdsixx) = 8.536 MeV and $S_n$(\gdsixx) = 7.937 MeV).  Since the fission channel is absent, and cross sections for charged-particle channels are very small, all $G^{CN}_{\gamma}(E,J,\pi)$ equal one below the neutron separation energy; their behavior above $S_n$ is governed by the competition of $\gamma$-decay and neutron evaporation.
We observe a dependence on the spin of the decaying nucleus that is stronger than in the actinide case, but significantly less than that found for the $^{92}$Zr example studied in Ref.~\cite{Forssen:07}; a parity dependence is also visible.  For energies below about 1 MeV, the branching ratios show effects of discrete levels in the neighboring nuclei; above that energy, the $G^{CN}_{\gamma}(E,J,\pi)$ have a smooth energy dependence.
While $\gamma$-branching ratios associated with small angular-momentum values ($J \leq 3$) are seen to drop rapidly right above the neutron separation energies, those related to larger $J$ values remain high ($G^{CN}_{\gamma}(E,J,\pi) = 1$) for several hundreds of keV above the neutron threshold.  For these higher-$J$ states, neutron evaporation is hindered at low energies, where s-wave neutron transmission dominates, since the residual  $^{155,157}$Gd nuclei contain few high-spin states to which the decay could occur.   With increasing excitation energy, states with higher spins become available in the neighboring nuclei and p-wave and d-wave transmission begin to compete -- neutron evaporation becomes the dominant decay mode.

As the excitation energy increases above values that correspond to $E_n \approx$ 1.5 MeV, the $G^{CN}_{\gamma}(E,J,\pi)$ begin to converge slowly.  However, within the energy range considered ($E_n = 0-3$ MeV), no particular energy region can be identified for which the \we limit is clearly reached.  

\begin{figure*}[htb]
\centering
\resizebox{0.8\columnwidth}{!}{
    \includegraphics[viewport=55 0 555 780,clip,angle=0]{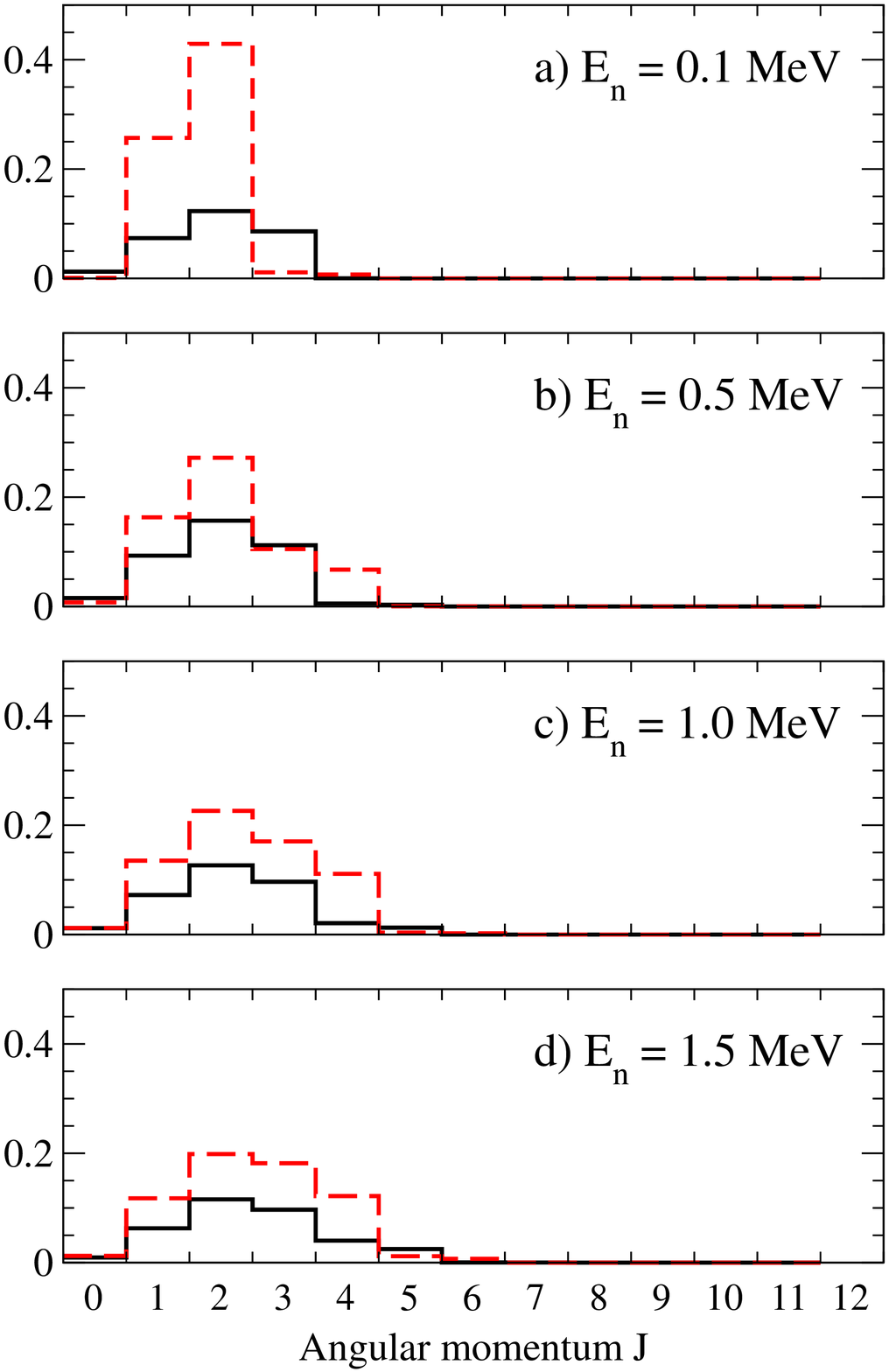}
}
\resizebox{0.8\columnwidth}{!}{
    \includegraphics[viewport=55 0 555 780,clip,angle=0]{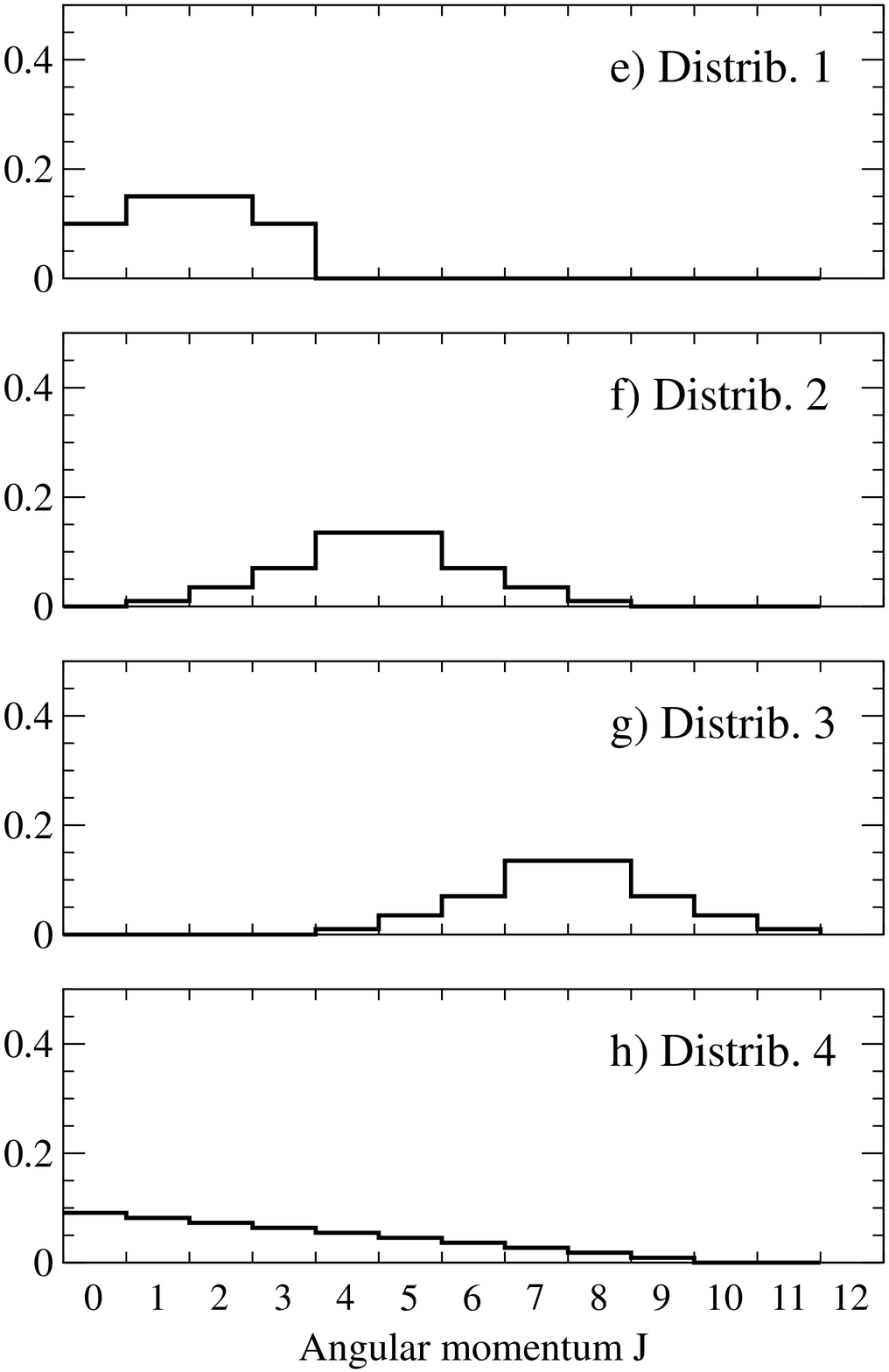}
}
\caption[Spin-parity distributions in $^{156}$Gd]{(Color online) Spin-parity distributions in $^{156}$Gd. 
Panels a) - d) show the distributions for the neutron-induced reaction for various neutron energies, from 0.1 to 1.5 MeV.  The positive-parity distribution is indicated with a solid line, the negative-parity distribution is given by a dashed line.  Panels e) - h) show schematic spin-parity distributions for simulated surrogate experiments.  Positive and negative parities are taken to be equally probable. 
}
\label{fig:JPiInGd}
\end{figure*}

\subsection{Weisskopf-Ewing approximation for \nga reactions in the rare-earth region}
\label{sec_rareearths_WE}

The spin-parity distributions relevant to neutron-induced reactions for \gdfive are shown in Fig.~\ref{fig:JPiInGd}a) -- d).  We find little to no contribution for angular momenta above $J=$5-6. Since the neutron-induced reaction produces only a relatively small range of angular-momentum values, one might expect the \we approximation to be reasonable, at least for $E_n >$ 0.5 MeV, where the $G^{CN}_{\gamma}(E,J,\pi)$ show a smooth behavior.  
To test this, we considered the four schematic surrogate spin-parity distributions shown in Fig.~\ref{fig:JPiInGd}e) -- f).  Positive and negative parities were taken to be equally probable for these cases, $p = 1,2,3,4$.  In addition, we considered the spin and parity-dependent distribution $p$ = ABB discussed earlier (Fig.~\ref{fig_jdistrib_uran}a).
We calculated simulated surrogate coincidence probabilities $P^{(p)}_{\delta\gamma}(E)$ for all five distributions and determined the related \we cross sections.  The results are shown in Fig.~\ref{fig:UngSimulWE_Gd}, where the extracted cross sections, for $p= 1,2,3,4,$ ABB, are compared to the \nga reference cross sections that were obtained from fits to direct measurements.  

For $E_n <$ 1.5 MeV, the cross sections extracted from a \we analysis of the simulated surrogate data are consistently too high, up to a factor of four for \gdfivex\nga and up to a factor of nine for \gdsevenx\ngax. These discrepancies are larger than in the actinide case, but smaller than what was observed for the zirconium region~\cite{Forssen:07}.
As expected, the largest deviations occur for $p=3$, \idest for the distribution that has the least overlap with the \jpi population of the compound nucleus in the neutron-induced reaction.  Distributions that peak at low spins, such as $p=1$ or ABB, yield much closer agreement with the reference cross section. 

For neutron energies above about 1.5 MeV, the cross sections for $p = 1,2,3,4$, and ABB begin to converge to the reference result, {\it i.e.\ } the \we approach becomes a better approximation.  In this region, most results agree with the reference cross section within about 10\%; only distribution 3, which contains contributions from angular momenta up to $J=11$, leads to larger deviations.
 
The results for \gdsevenx\nga are qualitatively similar to those for  \gdfivex\ngax, but differ in some crucial details.  In particular, the \we approach overestimates the \nga cross section by factors that are larger than those for the \gdfivex\nga case.  

\begin{figure}[htb]
\vspace{0.5cm}
\centering
\resizebox{1.0\columnwidth}{!}{
    \includegraphics[angle=270,viewport=80 15 530 730,clip]{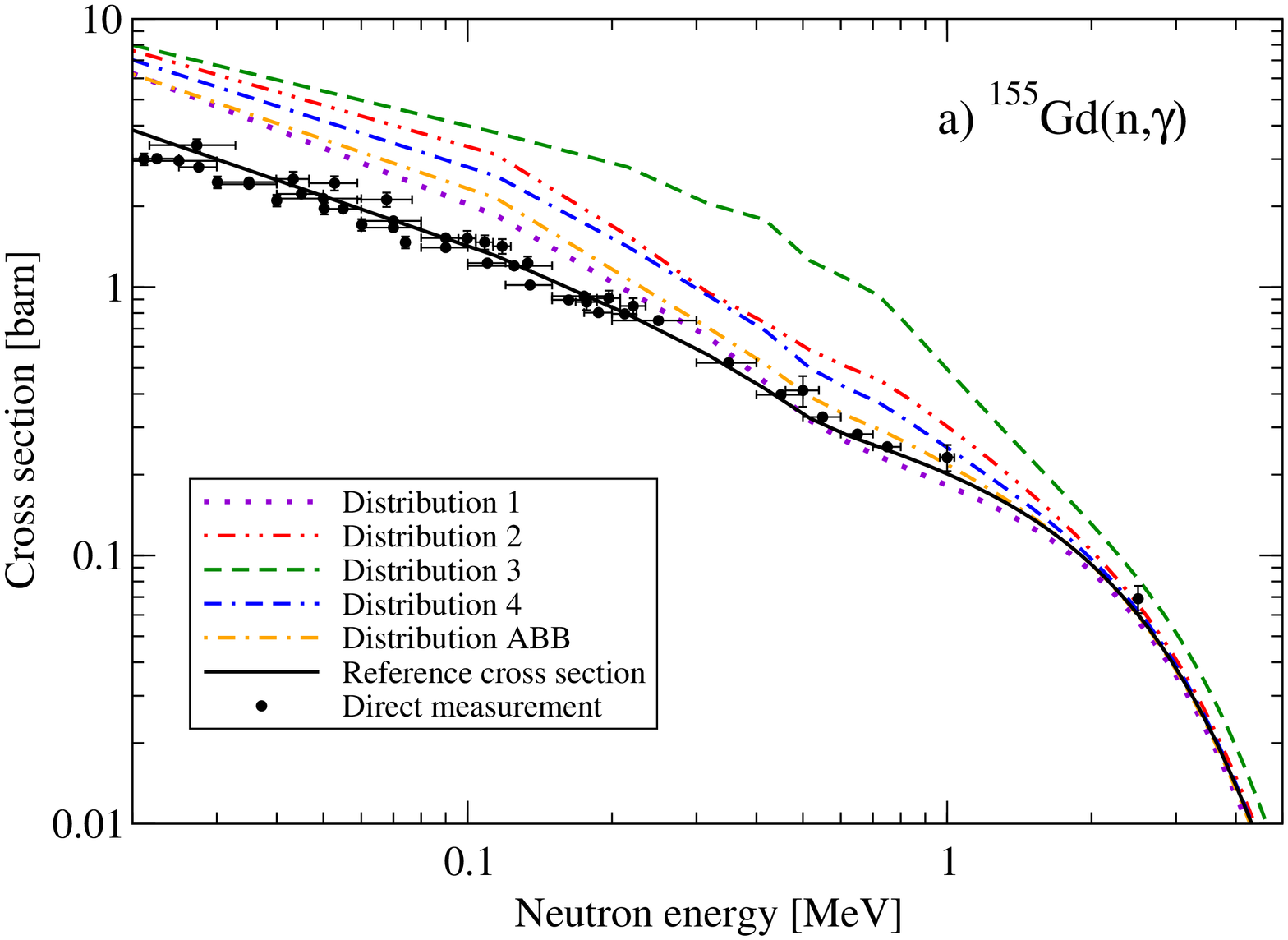}
}
\hfill
\resizebox{1.0\columnwidth}{!}{
    \includegraphics[angle=270,viewport=80 15 590 730,clip]{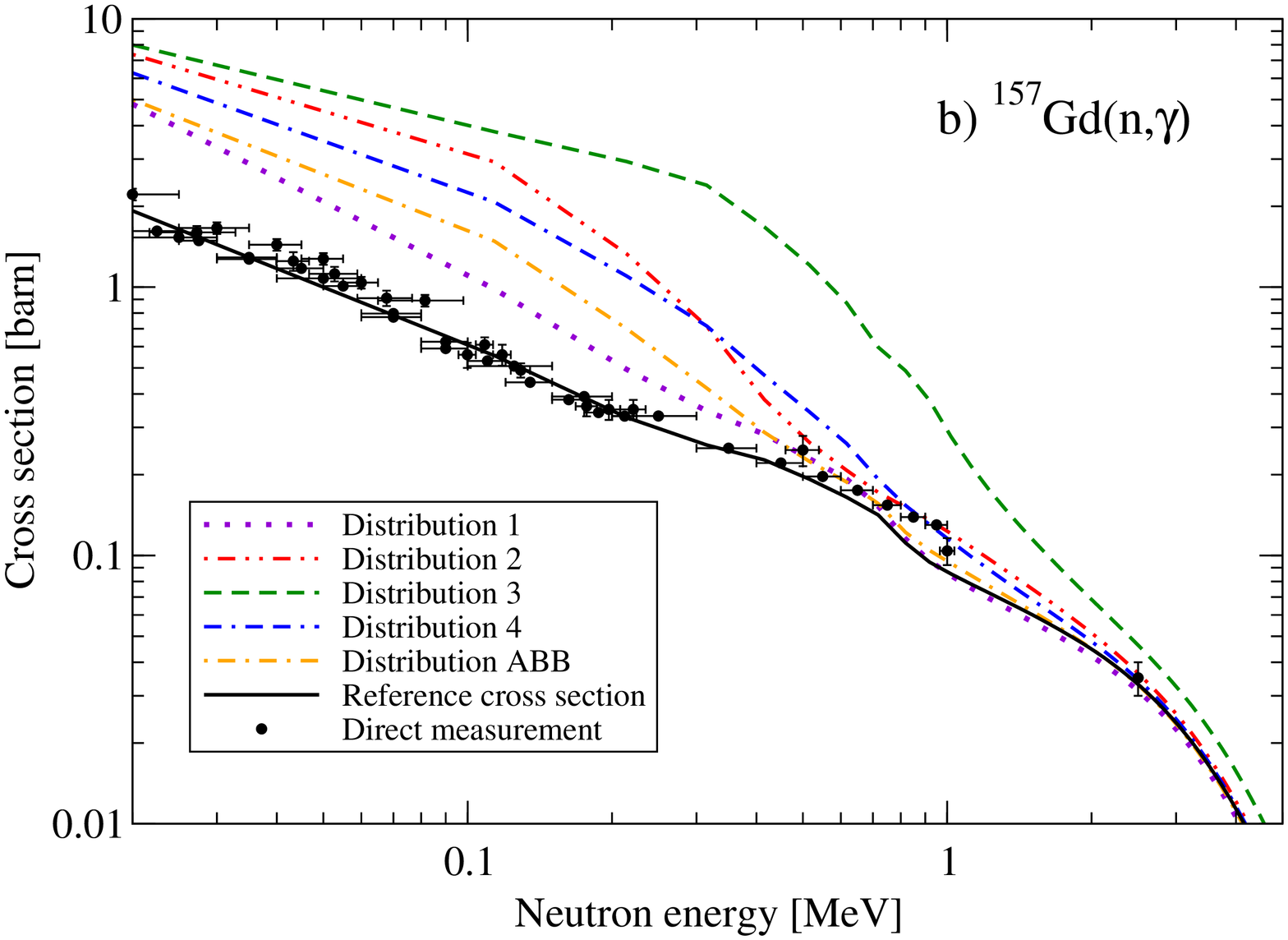}
}
\caption[Weisskopf-Ewing estimates for the a) $^{155}$Gd(n,$\gamma$) and b) $^{157}$Gd(n,$\gamma$) cross sections from simulated surrogate experiments]
{(Color online) Weisskopf-Ewing estimates for the a) $^{155}$Gd(n,$\gamma$) and b) $^{157}$Gd(n,$\gamma$) cross sections, extracted from analyses of simulated Surrogate experiments, for five different compound-nuclear \jpi distributions. For comparison, the reference cross section, which was obtained by adjusting the parameters for the Hauser-Feshbach calculation to reproduce direct measurements and evaluated results, is shown as well.
}
\label{fig:UngSimulWE_Gd}
\end{figure}

\subsection{Ratio approach for \nga reactions in the rare-earth region}
\label{sec_rareearths_ratios}

Since the factors by which the \we approach overestimates the \nga cross sections for \gdfive and \gdseven are different from each other, we expect that an external surrogate ratio analysis will not resolve the discrepancies.  To see whether the ratio approach at least reduces the deviations, we compare, in Fig.~\ref{fig:UngSimulRatio_Gd}, the ratios of our simulated coincidence probabilities, $R^{(p)}(E) =$ $P^{158Gd (p)}_{\delta,\gamma} (E)$/$P^{156Gd (p)}_{\delta,\gamma} (E)$, with $P^{(p)}_{\delta,\gamma}$ as defined in Eq.~\ref{eq:simRatio}, to the ratio of the reference cross sections, $\sigma$[\gdsevenx\ngax] /  $\sigma$[\gdfivex\ngax].  The simulated data result in ratios that differ from the reference ratio by as much as 250\% for energies below about $E_n =$ 0.7 MeV. The result for distribution 3 shows the largest deviations; ratios associated with distributions 1 and ABB differ by 20-50\% from the reference values. For energies above 0.7 MeV, all ratios converge toward the reference result.  At 1 MeV, the largest deviation is 35\% (for distribution 3) and most results lie within 10\% of the expected value.

Overall, the ratio approach reduces, but does not eliminate, the effect of the spin-parity mismatch on the extracted cross sections for energies where the Weisskopf-Ewing approximation is not valid.

\begin{figure}[htb]
\vspace{0.5cm}
\centering
\resizebox{1.0\columnwidth}{!}{
    \includegraphics[angle=270,viewport=80 35 590 730,clip]{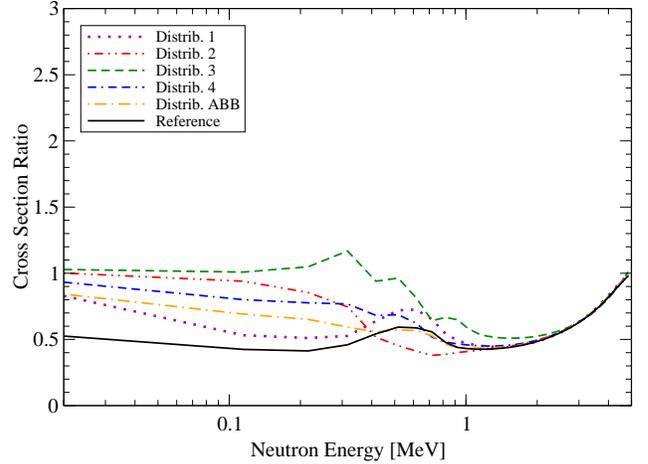}
}
\caption[Ratio results for the $^{157}$Gd(n,$\gamma$) to $^{155}$Gd(n,$\gamma$) cross sections from simulated Surrogate experiments]
{(Color online) Ratio results for the gadolinium nuclei. The ratio of the simulated surrogate coincidence probabilities $R^{(p)} =$ $P^{158Gd (p)}_{\delta,\gamma}$/$P^{156Gd (p)}_{\delta,\gamma}$, for five schematic \jpi distributions, is compared to the ratio of the reference cross sections.
}
\label{fig:UngSimulRatio_Gd}
\end{figure}


\section{Conclusions}
\label{sec_conclusions}

Motivated by the renewed interest in the surrogate nuclear reactions approach, we have examined the prospects for determining \nga cross sections for deformed rare-earth and actinide nuclei from surrogate measurements. In particular, we investigated the validity of approximation schemes that are commonly employed when extracting \nf cross sections from surrogate experiments.  The \we and ratio approaches, which neglect the fact that the spin-parity population of the compound nucleus produced in the surrogate reaction is different from that of the compound nucleus occurring in the desired reaction, were tested with calculations that simulated observables for typical surrogate experiments.  The approach used here is similar to the method employed in our earlier study of \nf reactions~\cite{EscherDietrich:06PRC} and complements and extends the investigation of \nga reactions for near-spherical nuclei in the mass 90-100 region~\cite{Forssen:07}.

Overall, we find that the probability for a compound nucleus to decay via $\gamma$ emission depends sensitively on the spin-parity population of the nucleus prior to decay.  The dependence of the $\gamma$-branching ratios on the \jpi distribution is greater than that found previously for fission.  Our studies of both gadolinium and uranium isotopes demonstrate that this is true in the presence as well as absence of an open fission channel, with the rare-earth nuclei exhibiting a somewhat stronger spin dependence than the actinide species studied.

\begin{figure*}[htb]
\centering
\resizebox{1.7\columnwidth}{!}{
    \includegraphics[viewport=60 25 620 780,clip,angle=270]{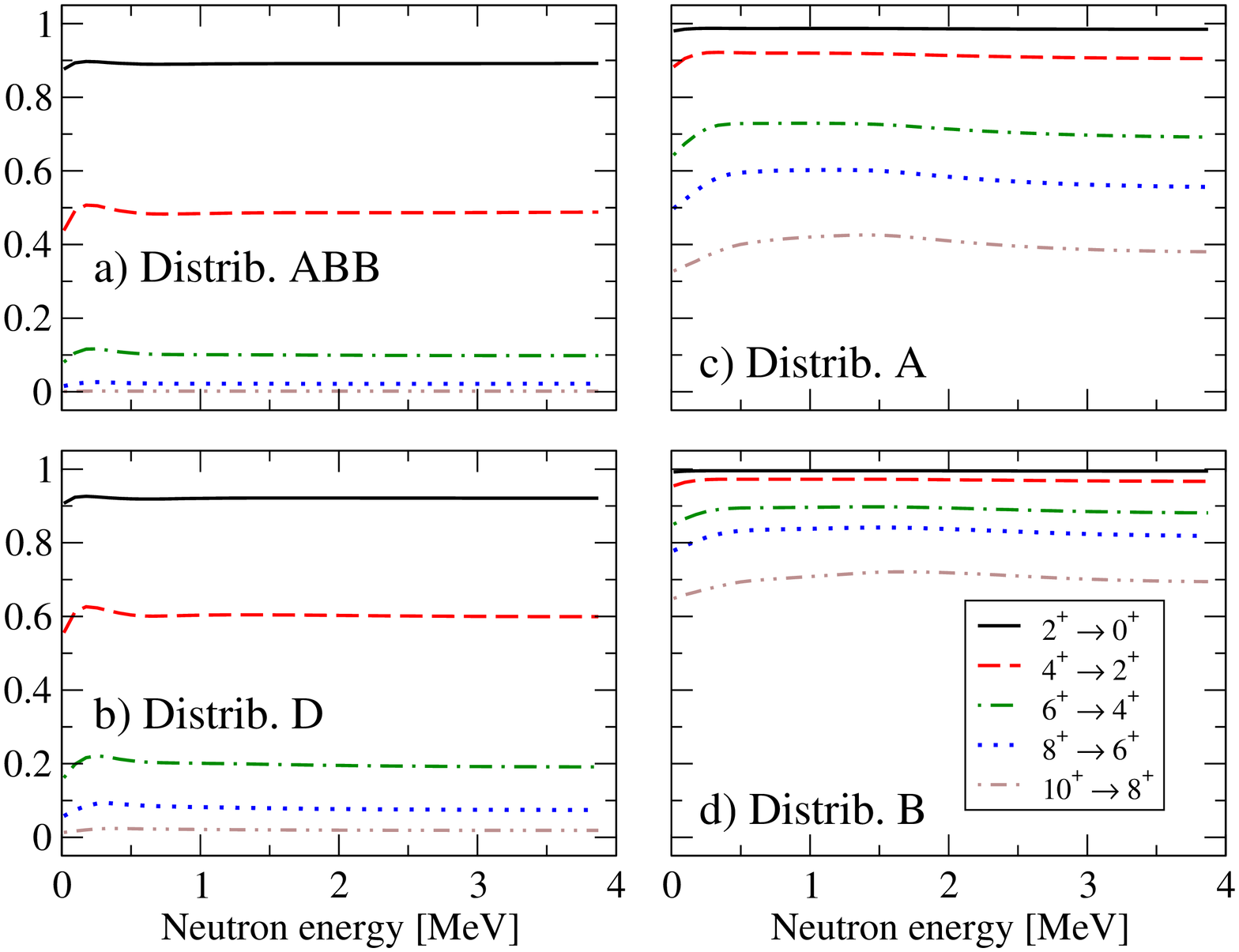}
    }
\caption[Gamma yields in \usix for different \jpi distributions]{Ratios of the yields of various $\gamma$-ray transitions in the ground-state band of $^{236}$U to the total production of $^{236}$U, for the four schematic spin distributions studied in Section~\ref{sec_actinides} and shown in Fig.~\ref{fig_jdistrib_uran}.}
\label{fig_236U_gYields}
\end{figure*}

We have shown that the \nga cross sections obtained when employing the \we or ratio approximations can differ significantly from the expected `true' cross section; for the cases considered here, we found deviations of a few percent to several hundred percent, depending on the severity of the spin-parity mismatch between the desired and surrogate reactions.  The uncertainty is clearly greater than that found for \nf cross sections and - unlike in  the fission case - is not substantially reduced when a ratio approach is used.   
At the same time, a comparison with the recent theoretical study of surrogate reactions in the zirconium region~\cite{Forssen:07}, which found a strong dependence of the $\gamma$-branching ratios on the spins of the compound nucleus, confirms the notion that the higher level densities present in the deformed rare-earth and actinide regions do indeed reduce the sensitivity of the $\gamma$-decay probabilities to compound-nuclear spin-parity distributions and nuclear-structure effects.

While it may be desirable to generally designate certain mass regions, energy regimes, or reaction types as good candidates for applying approximate methods in the analysis of surrogate data, the situation is quite subtle:  the validity of the \we approximation (and the related ratio methods) depends on the range of spins populated in both the desired and surrogate reactions.  For the cases studied here, namely low-energy ($E_n = 0-3$ MeV) \nga reactions on $^{155,157}$Gd and $^{233,235}$U targets, neither the \we nor the ratio method resulted in a satisfactory agreement of the extracted  cross section with the expected result (the known reference cross section), unless the compound-nuclear \jpi distribution produced in the surrogate reaction was taken to be similar to that found in the desired reaction.  Our findings are expected to be valid more generally for deformed rare-earth and actinide species.
Consequently, to obtain accurate \nga cross sections from surrogate measurements, it is necessary to design experiments in a manner that minimizes the spin-parity mismatch.  Alternatively, one can introduce theoretical techniques that account for the mismatch.  Obtaining information on the compound-nuclear \jpi distribution in the surrogate reaction is important for both approaches.

To properly predict the relevant spin-parity populations, a theoretical treatment of the processes that form the compound nucleus in a surrogate reaction is required.  This involves a description of direct reactions that populate highly-excited, unbound states, and the damping of these doorway states into more complicated configurations that lead to a compound nucleus.  The possibility that the intermediate system produced in a given surrogate reaction does not lead to the compound nucleus of interest, but decays via non-equilibrium particle emission prior to reaching the compound stage, has to be considered.  The probability for this process needs to be calculated, along with its dependence and influence on angular momentum, parity, and energy of the decaying nuclear system.  
Existing reaction-theory tools (DWBA and coupled-channels codes) can be modified to describe the first step of the surrogate reaction, the formation of a highly-excited intermediate system prior to equilibration.  Applications to various surrogate mechanisms have already been investigated~\cite{Andersen:70,Younes:03a,Younes:03b,Escher:07cnr}, and an initial study of the equilibration process is presented in Ref.~\cite{Dietrich:07cnr}.

Developing a reliable theoretical description of the formation of a compound nucleus following a direct reaction will be crucially important for improving the accuracy and reliability of the surrogate method and for extending its applicability beyond (n,f) reactions on actinide targets to other reaction types and mass regions.  Modeling the decay of the compound nucleus produced in a surrogate reaction should, in ideal circumstances, not be necessary, but will be very useful for developing and testing the surrogate approach.  For instance, modeling the $\gamma$ cascade will be necessary if individual $\gamma$ transitions are to be used to identify the exit channel of interest, since the yield associated with a particular transition is only a fraction of the quantity needed, the total yield associated with the sum of all $\gamma$-cascades. Alternatively, one can consider other experimental methods for identifying the $\gamma$ channel of interest, e.g. by employing calorimetric measurements of the $\gamma$-rays emitted in the decay.

While the strong spin-parity dependence of the observables used to tag the exit channel makes extracting \nga cross sections from surrogate measurements very challenging, it also provides valuable information. In particular, simultaneously measuring the yields of several $\gamma$-ray transitions of a decaying \cn can provide signatures for the spin-parity distribution of the \cn prior to decay.
An example for this is shown in Fig.~\ref{fig_236U_gYields}, where we have plotted the relative yields of several ground band transitions for \usixx, for the four schematic \jpi distributions shown in Fig.~\ref{fig_jdistrib_uran}.
We find that different \jpi distributions lead to markedly different relative $\gamma$-ray yields.  These observables can be employed to test and constrain theories that predict compound-nuclear spin-parity distributions.
Relative $\gamma$-ray yields for the decay of even-even gadolinium nuclei have recently been measured~\cite{Scielzo:09ip} and methods are being developed to use this information in order to improve the \nga cross sections determined from surrogate experiments.
This approach, if successful for \nga applications, will also help to improve the accuracy of low-energy fission cross sections extracted from surrogate experiments.


\begin{acknowledgments}
The authors appreciate constructive input from R. D. Hoffman and N. D. Scielzo.
This work was performed under the auspices of the U.S. Department of Energy by Lawrence Livermore National Laboratory under contract DE-AC52-07NA27344.
\end{acknowledgments}


\section*{Appendix: Optical-model potential for actinides}
\label{sec_opticalmodel}

The choice of optical model parameters is important for the theoretical
calculation of cross sections of interest.  The optical model enters into several aspects of our calculations:

\begin{itemize}
   \item It determines the cross section for formation of the compound nucleus in the initial neutron-target interaction;
   \item it is used to compute the transmission coefficients used in the Hauser-Feshbach calculations;
   \item it determines the cross sections for inelastic excitation of the coupled states in the ground-state rotational band.
\end{itemize}

The dependence of the elastic and direct non-elastic reaction cross sections on the optical model
parameters is detailed in the report~\cite{Chen:01}.  In the work in Ref.~\cite{Chen:01} a preliminary version (Flap~1.5) of a regional potential tuned for actinides was employed.  In Ref.~\cite{EscherDietrich:06PRC}, as well as in the present work, an improved potential (Flap~2.2) was used that was originally developed as part of the evaluation of the $^{239}$Pu(n,2n) cross section by a subtraction technique~\cite{Navratil:00}.  The parameters of both optical potentials are shown in Tables~\ref{table:flap15} and \ref{table:flap22}, and their predictions for total and compound cross sections are compared with the relevant experimental cross sections in Fig.~\ref{fig:ompresults}.  All calculations are carried out with the coupled-channel code {\sc ecis95}~\cite{Raynal:94}, using the option for relativistic kinematics. The potential is expected to be useful for neutron energies in the range of 0 to 60 MeV.

\begin{table}[htb]
\caption[Parameters for Flap~1.5 regional actinide optical potential]{Parameters for Flap~1.5 regional actinide optical potential.  The asymmetry parameter $\eta$ is $(N-Z)/A$, where $N,Z,A$ are the neutron, proton, and mass numbers of the target.  Energies are in MeV, and lengths in fm.}
\begin{center}
\begin{tabular}{l c}
\hline \hline
\multicolumn{2}{l}{\it Real Volume}         \\
\hspace{.1in} $V_{R}$   & $52.0-0.3E-(26.0-0.15E)\eta$ \\
\hspace{.1in} $r_V$     & 1.25  \\
\hspace{.1in} $a_V$     & 0.63  \\
\hline
\multicolumn{2}{l}{\it Imaginary Volume}    \\
\hspace{.1in} $W_{V}$   & $
\left\{ \begin{array}{l l}
0,                           & E \leq 10 \\
-3.8+0.38E-(-1.9+0.19E)\eta, & E > 10    \\
\end{array} \right. $ \\
\hspace{.1in} $r_W$     & 1.27  \\
\hspace{.1in} $a_W$     & 0.62  \\
\hline\hline
\multicolumn{2}{l}{\it Imaginary Surface}  \\
\hspace{.1in} $W_{S}$   & $
\left\{ \begin{array}{l l}
3.08+0.4E-(1.54+0.2E)\eta,       & E \leq 10 \\
8.496-0.142E-(4.248-0.071E)\eta, & E > 10    \\
\end{array} \right. $ \\
\hspace{.1in} $r_S$     & 1.27  \\
\hspace{.1in} $a_S$     & 0.62  \\
\hline
\multicolumn{2}{l}{\it Real Spin Orbit}   \\
\hspace{.1in} $V_{so}$  & 6.2   \\
\hspace{.1in} $r_{so}$  & 1.15  \\
\hspace{.1in} $a_{so}$  & 0.75  \\
\hline \hline
\end{tabular}
\end{center}
\label{table:flap15}
\end{table}

\begin{table}[htb]
\caption[Parameters for Flap~2.2 regional actinide optical potential]{Parameters for Flap~2.2 regional actinide optical potential.  This is a piecewise-linear potential, so that parameters are to be interpolated linearly between the indicated energies.  The strength parameters are given in an isospin representation (subscript 0 for isoscalar, 1 for isovector), which are to be combined as $U=U_0-U_1\eta$, where $\eta$ is the asymmetry parameter $(N-Z)/A$.  Energies are in MeV, and lengths in fm.  The spin-orbit potential is the same as for Flap~1.5 (see Table~\ref{table:flap15}).}
\begin{center}
\begin{tabular}{l c c c c c c}
\hline \hline
Energy                  & 0      & 1      & 5      & 10     & 20     & 50     \\
\hline
\multicolumn{7}{l}{\it Real Volume}                                           \\
\hspace{.1in} $V_{R0}$  & 52.000 & 52.000 & 51.661 & 49.856 & 46.810 & 38.351 \\
\hspace{.1in} $V_{R1}$  & 26.000 & 26.000 & 25.830 & 24.928 & 23.405 & 19.175 \\
\hspace{.1in} $r_V$     & 1.250  & 1.249  & 1.245  & 1.240  & 1.230  & 1.210  \\
\hspace{.1in} $a_V$     & 0.63   & 0.63   & 0.63   & 0.63   & 0.63   & 0.63   \\
\hline
\multicolumn{7}{l}{\it Imaginary Volume}                                      \\
\hspace{.1in} $W_{V0}$  & 0.000  & 0.000  & 0.000  & 0.338  & 2.143  & 7.557  \\
\hspace{.1in} $W_{V1}$  & 0.000  & 0.000  & 0.000  & 0.169  & 1.072  & 3.779  \\
\hspace{.1in} $r_W$     & 1.270  & 1.270  & 1.270  & 1.270  & 1.270  & 1.270  \\
\hspace{.1in} $a_W$     & 0.62   & 0.62   & 0.62   & 0.62   & 0.62   & 0.62   \\
\hline\hline
\multicolumn{7}{l}{\it Imaginary Surface}                                     \\
\hspace{.1in} $W_{S0}$  & 3.080  & 3.480  & 4.737  & 6.768  & 6.768  & 1.354  \\
\hspace{.1in} $W_{S1}$  & 1.540  & 1.740  & 2.369  & 3.384  & 3.384  & 0.677  \\
\hspace{.1in} $r_S$     & 1.270  & 1.270  & 1.270  & 1.270  & 1.270  & 1.270  \\
\hspace{.1in} $a_S$     & 0.62   & 0.62   & 0.62   & 0.62   & 0.62   & 0.62   \\
\hline \hline
\end{tabular}
\end{center}
\label{table:flap22}
\end{table}

\begin{figure}[htb]
\begin{center}
  \includegraphics[width=.5\textwidth]{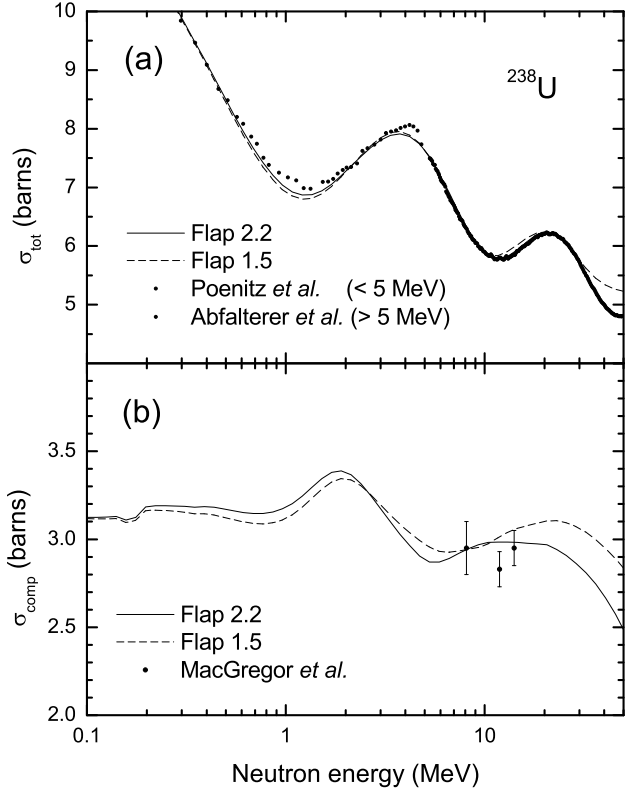}
\end{center}
  \caption[Comparison of optical-model total and compound cross sections with experiment for n + $^{238}$U]{Comparison of optical-model a) total and b) compound cross sections with experiment for neutrons incident on $^{238}$U.  The calculated compound cross section is the complete reaction (nonelastic) cross section with the inelastic cross sections for the ground-state rotational band removed.  The Flap~2.2 total cross section is indistinguishable from the experimental data above 5 MeV.}
  \label{fig:ompresults}
\end{figure}

The parameterization of the optical potential is a standard one (see, {\it e.g.}, Ref.~\cite{Becchetti:69}), employing Woods-Saxon volume form factors for the real and volume-imaginary potentials, a derivative Woods-Saxon for the surface-imaginary potential, and a Thomas form for the spin-orbit potential.   The strength and geometry parameters are shown in Tables~\ref{table:flap15} and~\ref{table:flap22}.  The Flap~2.2 potential is a piecewise-linear potential that allows an energy-dependent geometry for the real potential that is in accord with expectations based on dispersive phenomenological and microscopic folding optical models~\cite{Finlay:85,Jeukenne:77,Jeukenne:77a}.  This treatment allows an improved reproduction of total cross sections compared to the energy-independent geometry model of Flap~1.5, as shown in Fig.~\ref{fig:ompresults}.  The new potential was constrained to coincide with the older one at zero energy, so as to preserve the excellent reproduction of the low-energy resonance parameters ($S_0$, $S_1$, and $R'$) that was attained with Flap~1.5.  Both potentials were developed using mainly data on $n +$$^{238}$U and $n +$$^{232}$Th.

The experimental total cross section data shown in Fig.~\ref{fig:ompresults} are from Ref.~\cite{Poenitz:81} below 5~MeV and from~\cite{Abfalterer:01} above that energy.  The Flap~2.2 results are indistinguishable from the experimental data in the upper region;  the energy-dependent geometry was required to achieve this result.  There is significant scatter in the available experimental data on the compound cross section.  We have chosen to show one set which we judge to be reliable, that of Ref.~\cite{MacGregor:63}. The agreement with both calculations is good at the level of approximately~3\%.

Transmission coefficients used in all stages in the present $^{235}$U(n,2n) calculations were generated for neutrons incident on $^{234}$U using a coupled-channel model ({\sc ecis}) in which the 0$^+$, 2$^+$, and 4$^+$ members of the ground-state band were coupled in a rotational model.  
The resulting compound-formation cross section is shown in Fig.~\ref{fig_sigreac_fsd}.
The calculation used experimental values for the deformation parameters ($\beta_2=0.198$ and $\beta_4=0.057$) that are typical in this region of the actinides.  For $^{239}$Pu(n,2n), similar calculations were performed on $^{238}$Pu, changing only the energies of the coupled states.  Calculations of the same type on $^{238}$U were made for the cross section results shown in Fig.~\ref{fig:ompresults}.  The use of a common set of transmission coefficients for all nuclei in each reaction is reasonable, since the mass range is small.

\begin{figure}[htb]
\vspace{0.2cm}
\begin{center}
\includegraphics[width=0.9\columnwidth]{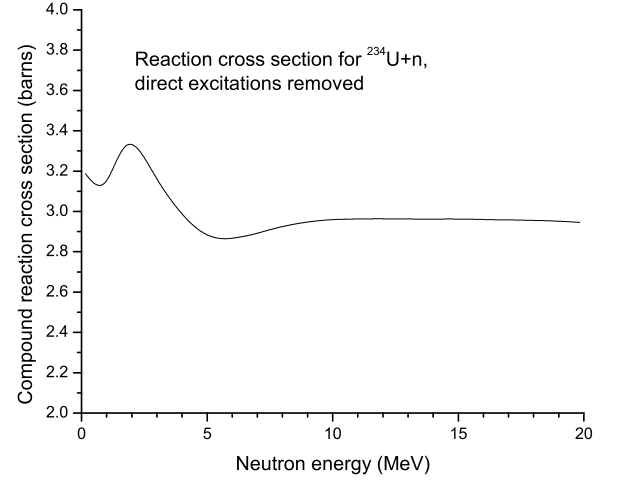}
\caption[Compound-nuclear reaction cross section for n+\ufour]{
Compound-nuclear reaction cross section for n+\ufourx.}
\label{fig_sigreac_fsd}
\end{center}
\vspace{0.1cm}
\end{figure}

Finally, we indicate the procedures used to obtain the transmission coefficients from the {\sc ecis} calculations.  Each channel in the coupled system is identified by quantum numbers $c \equiv [(ls)jI]J^{\pi}$, where the order of the symbols indicates a particular coupling scheme leading to total angular momentum and parity $J^{\pi}$ for the entire system.  In this scheme the relative orbital angular momentum $l$ is coupled to the projectile spin $s$ to a resultant $j$, which is then coupled to the target spin $I$ to yield the total angular momentum $J$.  The transmission coefficients are obtained from the calculated $S$-matrix elements by the well-known expression
\begin{equation}
T_c = 1 - \sum_{c'} \left\vert S_{cc'} \right\vert^2.
\label{eq:omp1}
\end{equation}
Whereas the transmission coefficients calculated from {\sc ecis} depend on the full set of quantum numbers $[(ls)jI]J^{\pi}$, those required by {\sc stapre} depend only on the orbital angular momentum $l$.  An averaging procedure is used to suppress the unwanted quantum numbers.  This procedure is rather arbitrary, but is chosen so that the most important quantity, the reaction cross section for compound nucleus formation, is preserved in the averaging procedure.  Following Ref.~\cite{Lagrange:83}, the dependence on total angular momentum $J$ is first removed,
\begin{equation}
T_{lsjI} = \sum_J \frac{2J+1}{(2I+1)(2j+1)} T^J_{lsjI},
\label{eq:omp2}
\end{equation}
then the dependence on $j$ is removed, 
\begin{equation}
T_{lsI} = \sum_j \frac{2j+1}{(2l+1)(2s+1)} T^J_{lsjI}.
\label{eq:omp2b}
\end{equation}
This is the desired expression in which the only variable is $l$, since $s$ and $I$ are fixed.  The transmission coefficients depend on the spin of the target state, $I$.  In practical calculations the target state is chosen as the lowest state of the ground-state rotational band.  In the present case we have used even-even targets so that $I=0$, and the above expressions simplify.  To our knowledge the accuracy of the averaging procedure and the appropriateness of using transmission coefficients based only on the ground state of the target have never been tested.  However, the preservation of the reaction cross section encourages the expectation that the errors incurred are not severe. 




\end{document}